\documentclass{iopjournal}

\usepackage{amsmath}
\usepackage{amssymb}
\usepackage{color,soul}

\providecommand{\DIFdel}[1]{} 

\begin{document}

\articletype{Article type} 

\title{Learning time-dependent and integro-differential collision operators from plasma phase space data using differentiable simulators}

\author{Diogo D. Carvalho$^{1,2,*}$\orcid{0000-0002-2663-3514}, Luís O. Silva$^{1}$\orcid{0000-0003-2906-924X} and E. Paulo Alves $^{2,3}$\orcid{0000-0002-4588-1003}}

\affil{$^1$GoLP/Instituto de Plasmas e Fus\~ao Nuclear, Instituto Superior T\'ecnico,
Universidade de Lisboa, 1049-001 Lisbon, Portugal}

\affil{$^2$Mani L. Bhaumik Institute for Theoretical Physics, University of California, California, Los Angeles, USA}

\affil{$^3$Department of Physics and Astronomy 
University of California, California, Los Angeles, USA}

\affil{$^*$Author to whom any correspondence should be addressed.}

\email{diogo.d.carvalho@tecnico.ulisboa.pt}


\begin{abstract}
Collisional and stochastic wave-particle dynamics in plasmas far from equilibrium are complex, temporally evolving, stochastic processes which are challenging to model. In this work, we extend previous methods coupling differentiable kinetic simulators and plasma phase space diagnostics to learn collision operators that account for time-varying background distributions. We also introduce a more general integro-differential operator formulation to probe relevant terms in the collision operator. To validate the proposed methodology we use data generated by self-consistent electromagnetic Particle-in-Cell simulations. We show that both approaches recover operators that can accurately reproduce the plasma phase space dynamics while being more accurate than estimates based on particle track statistics. These results further demonstrate the potential of using differentiable simulators to infer collision operators for scenarios where no closed form solution exists or deviations from existing theory are expected.
\end{abstract}

\section{Introduction}

Accurately capturing the effect of collisions in plasmas is a fundamental requisite to correctly model plasma dynamics over a broad range of densities and temperature conditions. Significant effort has been made to derive accurate theoretical descriptions of the statistical effect of collisions in plasmas~\cite{boltzmann1970weitere, landau1936kinetische, bhatnagar1954model, rosenbluth1957fokker, braams1987differential}. This description is commonly represented in the form of a collision operator $\mathcal{C}$, which acts on the velocity distribution function $f(\boldsymbol{v)}$ via:
\begin{equation}
    \left(\frac{\partial{f}}{\partial{t}}\right)_{col} = \mathcal{C}\left[f\right]
\end{equation}

For regimes where small angle scattering collisional dynamics dominate (non-strongly coupled plasmas), the collision operator is well described by a Fokker-Planck (FP) form~\cite{rosenbluth1957fokker}:
\begin{equation}
    \mathcal{C}[f]_{FP} = - \nabla_{\boldsymbol{v}}\boldsymbol{\cdot}(\boldsymbol{A}f) + \frac{1}{2}\nabla_{\boldsymbol{v}}\boldsymbol{\cdot}\left[\nabla_{\boldsymbol{v}}\boldsymbol{\cdot}(\boldsymbol{D}f)\right]
\label{eq:fp_operator}
\end{equation}
where $\boldsymbol{A} \in \mathbb{R}^N$ represents the advection (drag), and $\boldsymbol{D} \in \mathbb{R}^{N}\times\mathbb{R}^N$ the diffusion (spread) of the distribution function undergoes in the $N$-dimensional velocity space. FP type operators are ubiquitous in physics. In plasma physics, for example, they are used as statistical models for anomalous transport in plasma turbulence~\cite{escande2007can, isliker2017particle, isliker2017fractional}, and to model non-thermal particle acceleration in astrophysical scenarios~\cite{martins2009ion, wong2020first, wong2025energy}.

While the existing theory of collisions in plasmas is well established, there exist regimes where theoretical assumptions are expected to break down, such as in strongly coupled systems~\cite{baalrud2012transport, baalrud2014extending} or electromagnetic dominated scenarios~\cite{braams1987differential, pike2016transport}. First-principles simulations, such as Molecular Dynamics (MD) and Particle-in-Cell (PIC) simulations~\cite{hockney2021computer, birdsall2018plasma}, play a critical role in investigating collisional dynamics in these complex regimes and in guiding the development of new theoretical models. However, accurately and efficiently extracting collisional operators from such simulation data remains challenging.

In recent years, there have been significant efforts to develop data-driven tools to improve the modelling of collisions in plasmas. These are usually divided into two branches: to accelerate the computation of collisional effects (usually a known collision operator which is costly to calculate in numerical codes), and, more recently, to learn operators from self-consistent data (where no collisional operator is being used to generate the dynamics).

The first set of works is significantly more extensive. A variety of surrogate models has been trained to approximate the Landau-Fokker-Planck~\cite{miller2021encoder, lee2023oppinn, noh2025fpl}, Rosenbluth-Fokker-Planck~\cite{chung2023data}, and Boltzmann~\cite{xiao2021using, xiao2023relaxnet, holloway2021acceleration, miller2022neural} collision operators. Physics-Informed Neural Networks (PINNs)~\cite{raissi2019physics} have also been extensively used to model equations that contain Fokker-Planck-type operators. This includes works on the Vlasov-Fokker-Planck~\cite{hwang2020trend}, the Vlasov-Poisson-Fokker-Planck~\cite{lee2021model}, and the Landau-Fokker-Planck~\cite{chung2023data} equations. Similarly, PINNs were used to solve the Boltzmann equation in the context of weakly ionized plasmas~\cite{zhong2022low, kawaguchi2022physics}, the simplified Boltzmann-BGK equation~\cite{lou2021physics, li2024solving, oh2025separable}, and relativistic Fokker-Planck equations which describe runaway electron dynamics in magnetic confinement fusion devices~\cite{mcdevitt2023physics, mcdevitt2025efficient, mcdevitt2025physics}.

The second, recently emerging, set of works includes using PINNs to learn FP operators that describe electron dynamics observed in Earth's radiation belts~\cite{camporeale2022data}, Graph Neural Networks (GNNs) that learn to capture the self-consistent collisional dynamics of a one-dimensional plasma with arbitrary degrees of collisionality~\cite{carvalho2024learning}, neural networks (NNs) that learn the effective forces between macroscopic charged particles in dusty plasma from experimental trajectories~\cite{yu2025physics}, data-driven collision operators learned from molecular dynamics (MD) data \cite{zhao2025data, zhao2025fast}, and FP operators which describe finite-size particle collisions in Particle-in-Cell (PIC) simulations \cite{carvalho2026learning}. We believe that this second, less explored, line of research has a significant potential to enable theoretical developments for collisional theory, and other relevant sub-domains, and to also provide interpretable models to be used in reduced plasma descriptions.

In this work, we focus on extending the methods proposed in~\cite{carvalho2026learning} to account for: 1) time-varying FP operators; 2) a more general integro-differential formulation (of which FP operators are a subset), therefore allowing for the discovery of deviations from a FP description. We conjecture that these efforts will be of use in the near future to study non-thermal particle acceleration scenarios and electromagnetically dominated collisional regimes by incorporating findings based on the techniques described here, in reduced theoretical or simulation models. In particular, generalized data-driven collision operators~\cite{zhao2025data, carvalho2026learning} can be integrated into existing numerical codes, while time-dependent operators can be used for post-hoc analysis of experimental~\cite{bergeson2025experimental}, observational~\cite{camporeale2022data}, and simulation data~\cite{wong2020first, wong2025energy}. We anticipate that these techniques will be crucial to study collisional and stochastic wave-particle dynamics in relativistic plasmas far from equilibrium such as those expected to be observed, e.g., in the laboratory for Inertial Confinement Fusion (ICF) experiments~\cite{atzeni2004physics}, or in astrophysical scenarios such as relativistic magnetized jets~\cite{alves2018efficient}, relativistic magnetized turbulence~\cite{wong2020first, wong2025energy, lemoine2022first, lemoine2023particle}, and shocks~\cite{martins2009ion, gargate2011ion}.

\section{Methods}

In this work, we generalize the core strategy proposed in~\cite{carvalho2026learning} originally used to learn time-invariant FP collision operators from plasma phase space data. We frame the inverse problem of finding the operator $\mathcal{C}$ that describes the phase space dynamics as an optimisation task:
\begin{equation}
    \min_{\boldsymbol{\theta}} \sum_i^{N} \mathcal{L}\left(\hat{f}^{(t+i)}\left(\mathcal{C}(\boldsymbol{\theta}), f^{(t)}\right) - f^{(t+i)}\right)
\label{eq:loss}
\end{equation}
where $\boldsymbol{\theta}$ are the free parameters of the operator, $\hat{f}^{(t+i)}$ is the predicted distribution function at time-step $t+i$ when evolving the dynamics the initial distribution function $f^{(t)}$ over $i$ time-steps using the operator $\mathcal{C}$, and $\mathcal{L}$ a scalar loss function which quantifies the error of the predicted dynamics (in our case the mean absolute error, MAE, over the phase space).

\begin{figure}
    \centering
    \includegraphics[width=0.9\linewidth]{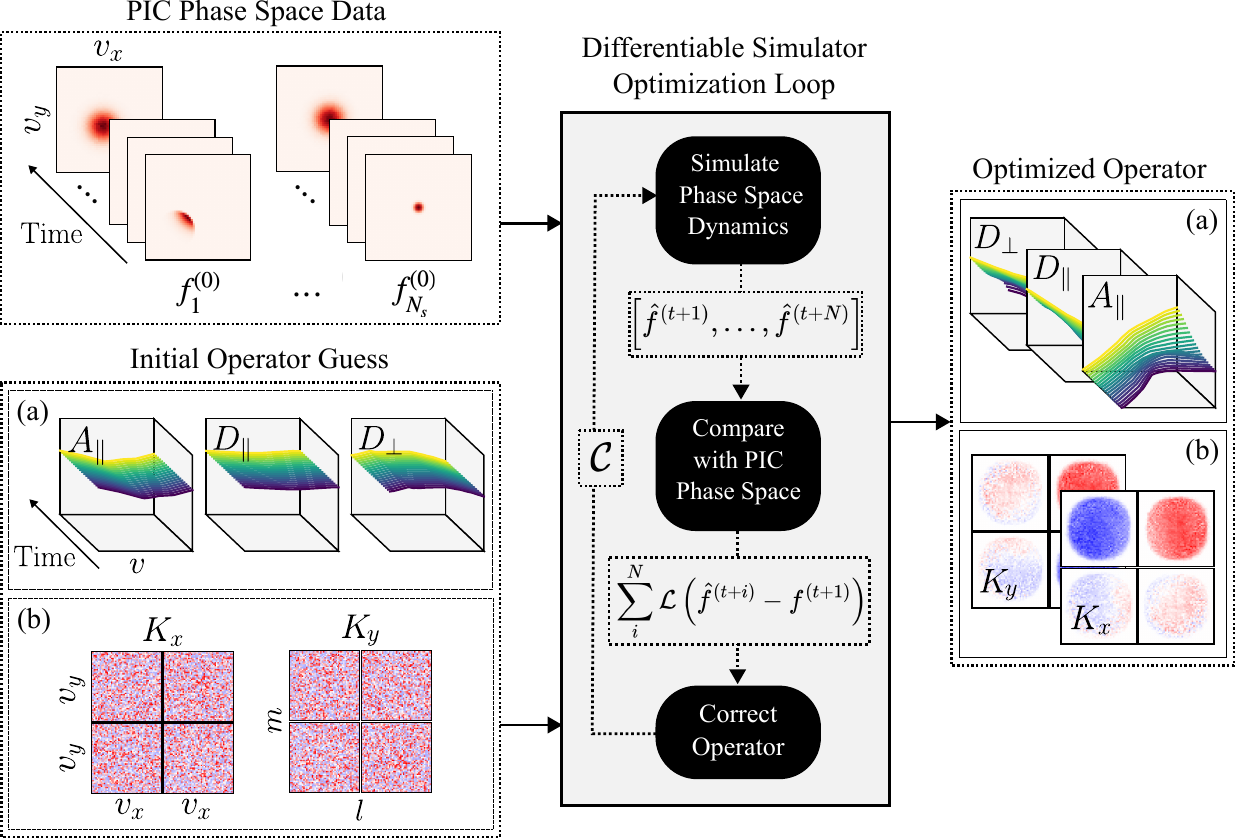}
    \caption{Illustration of the methodology used to learn collision operators from 2D PIC simulation data. Similarly to~\cite{carvalho2026learning} we use a differentiable simulator, coupled with phase space diagnostics from subpopulations of particles of the background plasma, to learn the collision operator ($\mathcal{C}$) that best describes the observed long-term phase space dynamics (i.e. that one that minimizes the loss $\mathcal{L}$~\eqref{eq:loss} between predicted $\hat{f}$ and observed $f$ distribution functions). In this work we explore two new description for the collision operator: (a) a Fokker-Planck time-dependent operator as described by \eqref{eq:fp_operator}, here represented solely by its parallel ($A_\parallel$, $D_\parallel$) and perpendicular ($D_\perp$) components with respect to the direction of propagation of the particle; (b) an integro-differential operator as described in \eqref{eq:K_operator} which in the discrete limit can be represented by a convolution of the distribution function with a 4-dimensional kernel $K_i(v_x,v_i,l,m)$ where the kernel sizes $k$ controls the non-locality of the  operator (in this case we show $k=2$). In both cases, the operators can be approximated either by a discrete tensor or a continuous approximator such as a NN. For the NN scenario (example in (a)), initial guesses correspond to randomly initialized models. For the discrete tensor scenario (example in (b)) the initial operator guess is set to zero (for illustrative purposes we instead set random values in this figure).}
    \label{fig:fp_solver_illustration}
\end{figure}

We tackle this optimisation task by using a differentiable simulator (implemented using PyTorch~\cite{ansel2024pytorch}) which includes two different families of collision operators: time-dependent FP operators as later defined in Section~\ref{methods:time_dependent_advection_diffusion_operator}; integro-differential operator as later defined in Section~\ref{methods:K}. The differentiable nature of the simulator enables the usage of a gradient-based optimisation algorithm (Adam~\cite{kingma2014adam}) to find the optimal values of $\theta$. To constrain the ill-posed problem, we optimise the values of $\theta$ over multiple initial subpopulations of particles and time-steps, as detailed in~\cite{carvalho2026learning}. An illustration of the overall methodology is provided in Figure~\ref{fig:fp_solver_illustration}. 

For all the implemented operators, we advance the dynamics in time using a forward Euler method, which we find to be stable for the problems under study and the time-steps used. During training we use a curriculum learning procedure which steadily increases the horizon over which forward dynamics are computed. As demonstrated in~\cite{carvalho2026learning}, this procedure enables us to learn improved estimates of the operator and consequently reduces the error in longer-term test rollouts. More details regarding the training procedure can be found in~\ref{app:training_procedure}.

\subsection{Time-Dependent Advection-Diffusion Operator}
\label{methods:time_dependent_advection_diffusion_operator}
When learning a time-dependent advection-diffusion operator, the operator form is given by eq.~\eqref{eq:fp_operator} where $\boldsymbol{A}(\boldsymbol{v}, t): \mathbb{R}^3 \rightarrow \mathbb{R}^2$ and $\boldsymbol{D}(\boldsymbol{v}, t): \mathbb{R}^3 \rightarrow \mathbb{R}^2\times\mathbb{R}^2$ are the functions to learn. We can approximate these functions using a continuous or discrete description. 

In the continuous case, we parameterise the functions using neural networks (NNs) such that $A_i(\boldsymbol{v},t) = NN_{A_i}(\boldsymbol{v},t)$ and $D_{ij}(\boldsymbol{v},t) = NN_{D_i}(\boldsymbol{v},t)$. The learnable parameters $\boldsymbol{\theta}$ correspond to the free parameters of the NN models.

In the discrete case, we consider that the operators are defined over a finite grid, e.g. $A_i \in \mathbb{R}^{N_v} \times \mathbb{R}^{N_v} \times \mathbb{R}^{N_t}$, where $N_v$ is the number of grid points in velocity space (assumed to be the same along $v_x$ and $v_y$) and $N_t$ the number of grid points along $t$. In this case, the learnable parameters are the values of the tensors. Note that the operator velocity and temporal grids do not have to match those of the simulation since they can always be interpolated to the desired position. Further smoothness can be softly enforced by adding a regularizing term in the loss function which represents a norm of the derivatives of the operator.

Finally, it is possible to enforce certain physics-inspired symmetries by using the strategies proposed in~\cite{carvalho2026learning}. For example, for an isotropic distribution function we expect that $\boldsymbol{A}(\boldsymbol{v},t) \equiv A_\parallel(v, t) \hat{\boldsymbol{v}}$ and $\boldsymbol{D}(\boldsymbol{v},t) \equiv D_\parallel(v,t)\hat{\boldsymbol{v}}\hat{\boldsymbol{v}}^T + D_\perp(v,t)(\mathbf{I} -\hat{\boldsymbol{v}}\hat{\boldsymbol{v}}^T)$ where $\hat{\boldsymbol{v}} = \boldsymbol{v}/v$, and $\mathbf{I}$ is the identity matrix (of order 2). Enforcing these symmetries significantly reduces the dimensionality of the inverse problem and makes it easier to learn a stable and smooth operator~\cite{carvalho2026learning}.

\subsection{Integro-Differential Operator}
\label{methods:K}
Advection-diffusion models are part of a larger family of non-local integro-differential operators which can be written as:
\begin{align}
    \mathcal{C}[f]_{ID} &= \nabla_{\boldsymbol{v}} \boldsymbol{\cdot} \left(\boldsymbol{K} * f\right)(\boldsymbol{v}) \\
&=\frac{\partial}{\partial v_x} (K_x * f)(\boldsymbol{v}) + \frac{\partial}{\partial v_y} (K_y * f)(\boldsymbol{v})
\label{eq:K_operator}
\end{align}
where $(K_i*f)(\boldsymbol{v})$ = $\int_{\mathbb{R}^2} K_i(\boldsymbol{v}, \boldsymbol{v} - \boldsymbol{v}')f(\boldsymbol{v}')\boldsymbol{dv}'$ is a convolution operation between a kernel function $K_i$, and the distribution function $f$. In the discrete scenario, where $f(\boldsymbol{v}) \in \mathbb{R}^{N_v}\times\mathbb{R}^{N_v}$, we can control the non-locality of the operator by considering that $K_i(\boldsymbol{v}, \boldsymbol{v} - \boldsymbol{v}')$ is compactly supported, i.e. non-zero only in a bounded region of size $k\times k$ cells around $\boldsymbol{v}$. The (discrete) convolution is then defined, as a cross-correlation, as:
\begin{equation}
    (K_i * f)(\boldsymbol{v}) = \sum_{l=-n_{lower}}^{n_{upper}} \sum_{m=-n_{lower}}^{n_{upper}} K_i(\boldsymbol{v}, l, m) f(v_x + l \Delta_v, v_y + m \Delta_v)
\end{equation}
where $n_{lower} = \lfloor (k-1)/2 \rfloor$, $n_{upper} = \lfloor k/2 \rfloor$, and $\Delta_v$ is the phase space resolution (assumed to be equal along both directions).
The size $k$ controls which terms this generalised operator can represent. For example, assuming that $\partial_{v_i}$ in \eqref{eq:K_operator} is computed using a forward difference scheme, advection can be captured when $k=1$ (local operator), diffusion requires $k=2$ (non-local operator), and for non-local transport $k\geq4$.

For scenarios where the operator form is not known \textit{a priori}, using this generalised form can therefore be beneficial as a first step to identify the relevant terms: a Pareto-Curve analysis over $k$ will reveal the meaningful terms without having explicitly implemented them beforehand.

Similarly to the FP case, a discrete or continuous representation of $\boldsymbol{K}$ can be learned. For the discontinuous case, this corresponds to learning the values of two tensors $K_i \in \mathbb{R}^{N_v} \times \mathbb{R}^{N_v} \times \mathbb{R}^k \times \mathbb{R}^k$. For the continuous case, this corresponds to learning the free parameters of NNs such that $K_i(\boldsymbol{v}) = NN_{K_i}(\boldsymbol{v})$. It is also possible to impose symmetries in the learned kernels without making significant assumptions about the underlying operator form. For example, if we expect that dynamics along $v_x$ and $v_y$ to be symmetric, then we can enforce that $K_x(v_x,v_y,l,m) \equiv K_y(v_y, v_x, m, l)$.

\subsection{Particle-in-Cell Simulations}

For data generation we have performed 2D PIC simulations using the electromagnetic code OSIRIS~\cite{fonseca2002osiris}. Across all simulations we consider a single mobile electron species moving over a fixed neutralizing ion background. For all scenarios, we use a charge-conserving current deposition scheme~\cite{esirkepov2001exact}, a finite-difference time domain Yee solver for the electromagnetic fields~\cite{yee1966numerical}, and the standard Boris pusher~\cite{birdsall2018plasma} to advance particle momenta and position. The collisional interactions between finite-size electrons are self-consistently modelled by the PIC loop and no additional routines are used to enhance collisionality. Therefore, the operators retrieved in this work correspond to Coulomb collision operators modified for finite-size particles~\cite{langdon1970nonphysical, langdon1970theory,touati2022kinetic} as verified in previous work for thermal plasma setups~\cite{carvalho2026learning}. For the setups under study, the rate of collisionality is controlled by varying the number of particles per cell ($N_{ppc}$), shape function order ($m$), and initial velocity distribution function, while keeping other simulation parameters known to affect collisionality fixed (e.g. grid resolution $\Delta_x$)~\cite{carvalho2026learning, hockney1971measurements}.

For the time-dependent operator in Section~\ref{sec:results_time_dependent_operator} we will study the relaxation of an initial electron isotropic waterbag velocity distribution defined as:
\begin{equation}
    f(\boldsymbol{v}) = 
\begin{cases} 
\frac{n_{0}}{\pi v_w^2} & \text{if } v \le v_w \\
0 & \text{otherwise}
\end{cases}
\ ,
\label{eq:waterbag}
\end{equation}
where $n_0$ is the background number electron density, $\boldsymbol{v}\in\mathbb{R}^2$ and $v=||\boldsymbol{v}||$, towards a thermal distribution due the finite-size electron collisions.  We chose this setup since it represents a situation far from equilibrium where the background distribution evolves significantly throghout the simulation.
For the integral-differential operator in Section~\ref{sec:results_K} we will consider a simpler scenario, with a thermal distribution, for which the operator does not change over time. More details regarding the simulation parameters are given in \ref{app:pic_simulation_parameters}. 

For all simulations, we stored the raw data of all particles (approximately $10^6$ particles) at fixed time-intervals. This information is used to construct the phase space information of different subpopulations in post-processing as described in \ref{app:subpopulation_sampling} and build an estimate of the collision operator as from particle tracks as described in \ref{app:estimation_of_coefficients_from_tracks}. We only store the data of all particles to be able to compare the operators recovered from their tracks against those recovered from the phase space evolution of subpopulations. In future works, this memory-intensive step will not be required as the methodology proposed and tested here can be applied directly to the phase space diagnostics.

\section{Results}

\subsection{Time-Dependent Advection-Diffusion Operator}
\label{sec:results_time_dependent_operator}

For the study of a time-dependent advection-diffusion operator we use as test-bench the relaxation of an isotropic waterbag distribution to a normal distribution. Since the evolving background distribution is isotropic, the FP operator can be defined only along the directions parallel and perpendicular to the propagation of the particle, meaning we only need to learn $A_\parallel(v)$, $D_\parallel(v)$, and $D_\perp(v)$. 

\begin{figure}
    \centering
    \includegraphics[width=0.8\linewidth]{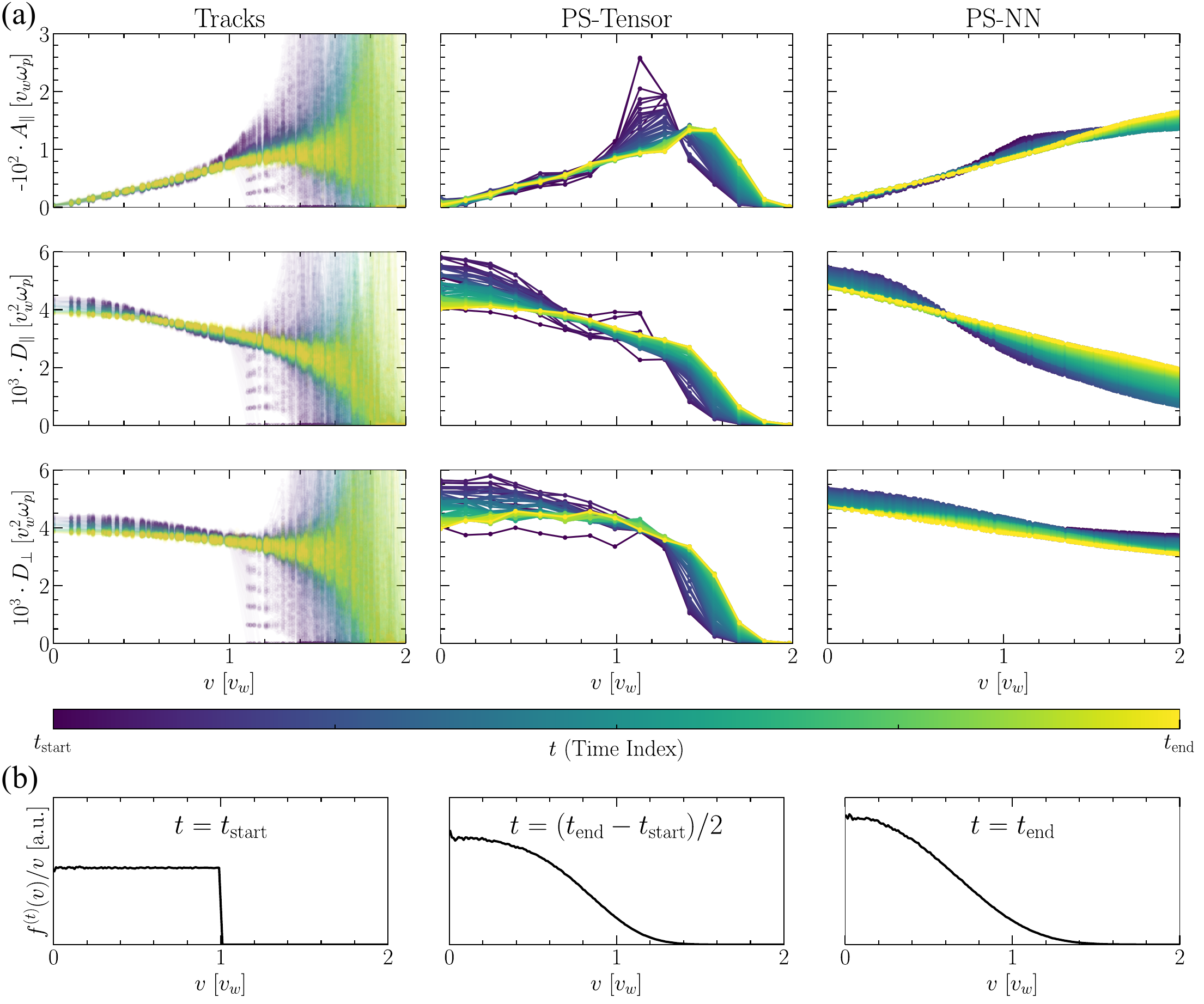}
    \caption{(a) Time-dependent advection-diffusion coefficients retrieved using  Tracks, and phase space based approaches with a discrete (PS-Tensor) and continuous (PS-NN) function approximators. Electron velocity distribution function is initialized using a waterbag distribution~\eqref{eq:waterbag} with radius $v_w = 0.02c$ which relaxes to a thermal distribution due to self-consistent, finite-size particle, electromagnetic collisional dynamics. Numerical simulation parameters (which affect collisionality) correspond to a number of particles per cell $N_{ppc}=4$, linear particle shape function, and grid resolution $\Delta_x = 0.01 c/\omega_p$ (more details in~\ref{app:pic_simulation_parameters}). Main difference to highlight is that the Diffusion predicted from Tracks is lower than the one measured from PS-Tensor and PS-NN. (b)~Snapshots of the velocity distribution $f(v) = v \int d\theta f(v\cos{(\theta), v\sin{(\theta))}}$ at different times. All curves are normalized to the same value. Initial velocity distribution function corresponds to an isotropic waterbag ($t=t_{start}$) which by the end of the simulation has relaxed towards a thermal distribution ($t=t_{end}$). Distribution function evolves more rapidly in the initial time-steps of the simulation. This is the reason why the advection-diffusion coefficients vary more noticeable for smaller $t$ values (purple curves).}
    \label{fig:AD-sim1-comparisons}
\end{figure}

In Figure~\ref{fig:AD-sim1-comparisons}(a) we showcase the operators retrieved via 3 methods: \textit{Tracks} -- estimated from particle tracks (according to the process described in \ref{app:estimation_of_coefficients_from_tracks}); \textit{PS-Tensor} -- using the phase space evolution of subpopulations and considering a discrete operator (as described in Section~\ref{methods:time_dependent_advection_diffusion_operator} using the 7 initial subpopulations listed in~\ref{app:subpopulation_sampling}); \textit{PS-NN} -- similar to PS-Tensor while parameterizing the operator with NNs. 

For the Tracks case, we observe a very smooth behaviour for regions with a large statistics (low $v$, large number of particles) while observing significant noise for higher $v$ (region with poor particle statistics). For the PS-Tensor case, curves are not as smooth, and the values obtained for the diffusion coefficients (specially at smaller $t$) are higher than those estimated from the tracks. The zero-valued entries at high $v$ are an artifact of the absence of statistics during training, and do not require further interpretation. Finally, for the PS-NN case we obtain smoother curves (given the inheritance smoothing introduced by the NN) which seem to be more in agreement with PS-Tensor case.

From Figure~\ref{fig:AD-sim1-comparisons}(b) we clearly observe that the background distribution function is evolving over time, starting as an initial waterbag distribution and relaxing towards a thermal distribution. This evolution of the background distribution function is what leads to time-dependent advection-diffusion coefficients.

To evaluate the accuracy of the different operators we produce rollouts for different initial subpopulations (i.e. simulate for each subpopulation the dynamics from $t=0$ up to $t=t_{max}$) and compute the average mean absolute error across the phase space, time-steps, and subpopulations ($\mathrm{MAE-Rollout}$). The values obtained are presented in Table~\ref{tab:sim-1-error}. 
\begin{table}
\centering
\caption{Average rollout error over multiple initial test subpopulations for operators retrieved in Figure~\ref{fig:AD-sim1-comparisons}. Error is significantly lower for phase space models indicating that measurement from Tracks provides a sub-optimal estimate.}
\label{tab:sim-1-error}
\begin{tabular}{lccc}
\hline
 & Tracks & PS-Tensor & PS-NN \\
\hline
$\mathrm{MAE-Rollout}$ $\times 10^2$  & $6.47 \pm 1.49$ & $3.34 \pm 0.20 $ & $3.34 \pm 0.16$ \\
\hline
\end{tabular}
\end{table}
It is clear that the operators learned using the phase space evolution of the subpopulations are more accurate than when estimated from particle tracks, thus confirming that the disagreements in the measured diffusion values are due to an incorrect estimate for the Tracks case.

\begin{figure}
    \centering
    \includegraphics[width=0.6\linewidth]{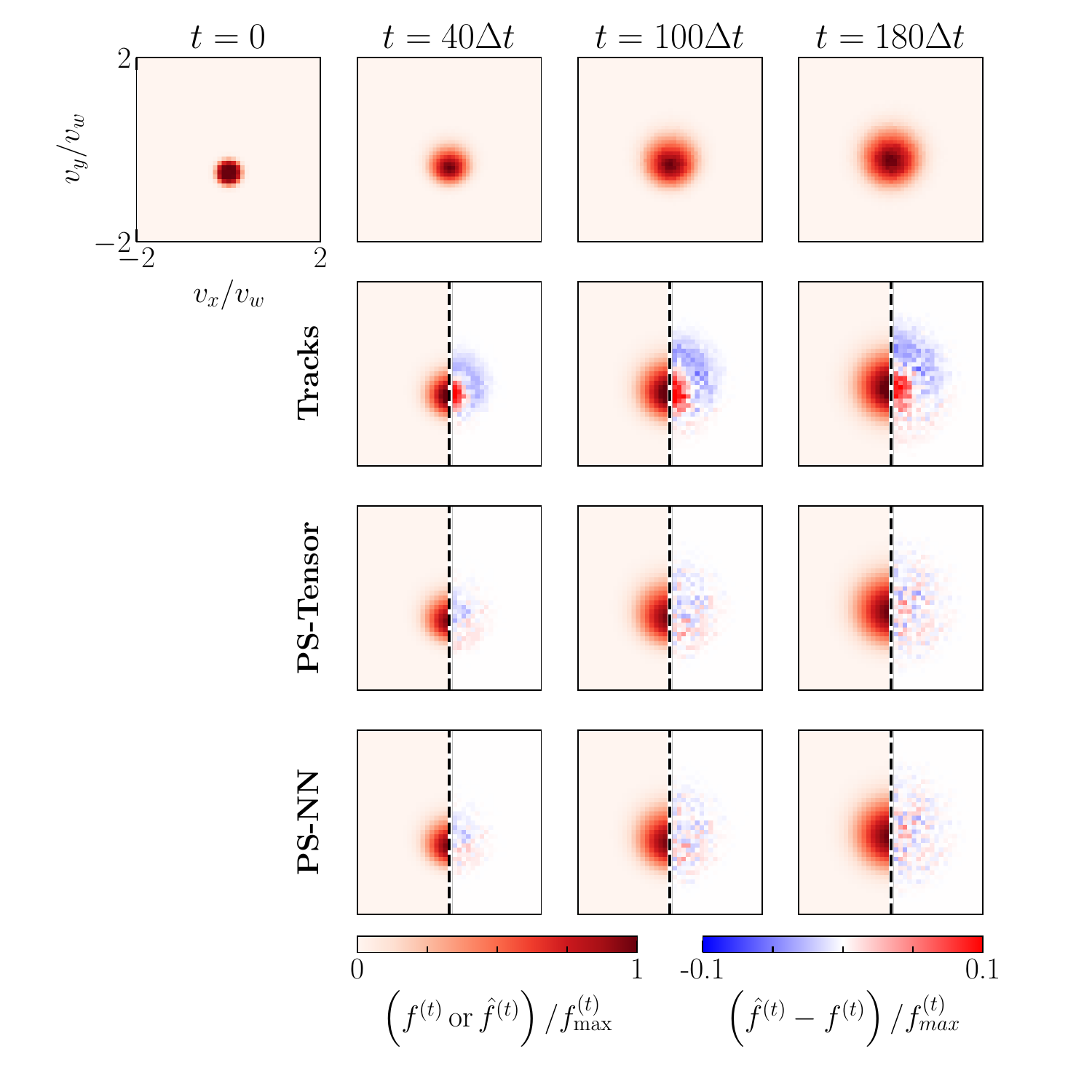}
    \caption{Phase space evolution for a  subpopulation using operators recovered in Figure~\ref{fig:AD-sim1-comparisons}. The top row corresponds to the observed dynamics in the PIC simulation ($f^{(t)}$). The remaining rows represent the predicted phase space evolution on the left ($\hat{f}^{(t)}$ for $v_x/v_{w}\in [-2,0]$) and the difference to the PIC data on the right ($\hat{f}^{(t)} - f^{(t)}$ for $v_x/v_{w}\in [0,2]$). Values are normalized to the peak of the PIC distribution function at time $t$ ($f^{(t)}_{max}$). The operator estimated from particle tracks fails to reproduce the phase space dynamics. The operators learned from phase space evolution using the differentiable simulator approximate the dynamics relatively well, and overall, the random distribution of errors can be attributed to the granularity of the original distribution function. Examples for other subpopulations are provided in~\ref{app:ad_extra_rollouts}.}
    \label{fig:AD-waterbag-ex_rolllout_dif_normal_-1_0}
\end{figure}

To illustrate the magnitude of the different errors we show an example of such rollouts and the corresponding predictions using the different methods in Figure~\ref{fig:AD-waterbag-ex_rolllout_dif_normal_-1_0}.
While the PS-Tensor and PS-NN method do not seem to produce systematic errors over a long rollout period, the operator retrieved from particle tracks does not correctly capture the phase space dynamics observed in the PIC data. We provide additional examples for different subpopulations, with similar conclusions, in~\ref{app:ad_extra_rollouts}.

It is important to clarify why estimates from particle tracks are incorrect for the simulation set under study: there is a lack of separation between the time-scales of the collisional processes and the collective plasma oscillations. The plasma period is being resolved by the phase space diagnostics ($\Delta t\omega_p \approx 1)$ in order to capture the smooth evolution of the distribution function. This should bias the particle tracks statistics since measures of $<\Delta v_i>$ and $<\Delta v_i\Delta v_j>$ over particles will contain the super-position of the e-e collisions, as well as the plasma oscillations. Since the PS-Tensor and PS-NN models learn the coefficients from the coarser phase space distribution, and are optimized for long-term accuracy, they do not seem to be impacted by the underlying oscillation. We further support this argument by repeating the analysis for an equivalent setup with lower collisionality (where collisional dynamics occur over larger time-scales than the plasma oscillation), see~\ref{app:estimation_of_coefficients_from_tracks} and~\ref{app:sim-2} . For that case, we observe a significant decrease in the errors obtained for estimates from tracks, while still observing a benefit in using the phase space based approach. These results are in agreement with similar findings for thermal plasma setups studied in~\cite{carvalho2026learning} and provide an important insight for future works who might want to use similar strategies to study strongly-coupled systems.

Finally, it is important to highlight that despite the PS-Tensor and PS-NN achieving similar test errors, their form is, nonetheless, different. This is related to the ill-posedness of the inverse problem which we are tackling when estimating advection-diffusion models from phase space data~\cite{carvalho2026learning}, i.e. there is a family of solutions which can produce statistically equivalent dynamics (i.e. equivalent rollout error). Our approach throughout this work is to constrain the space of solutions by reducing the degrees of freedom of the operator, use more (varied) subpopulations, and regularization methods. Overall, we find that increasing the number of training subpopulations and enforcing known physical symmetries are the most robust approaches to increasing the generalization capabilities of the operator and reduce the estimated coefficients variability (in regions of the phase space where dynamics are observed during training). Techniques which attempt to impose smoothness constraints via an extra loss term or by artificially reducing the degrees of freedom (e.g. along the time axis for PS-Tensor operator) can help to consistently recover smoother operators but do not address the non-uniqueness problem. The impact of all these choices is covered in more detail in~\ref{app:discussion_non_uniqueness}, \ref{app:regularization_vs_nt}, and previous work~\cite{carvalho2026learning}. We expect the non-uniqueness scenario to be a recurring challenge in future works, so it is important to further develop approaches to mitigate it, as proposed here. For example, we conjecture that by reducing the noise level and granularity of the phase space dynamics (by using a larger number of particles per subpopulation and a higher phase space resolution) we will be able to reduce the variability of the estimated coefficients.

\subsection{Integro-Differential Operator}
\label{sec:results_K}

A general benefit of the proposed approach, which combines phase space data with the differentiable simulator, is the significant flexibility to infer the form of the operator in post-processing due to reduced storage costs and the flexibility to parameterize the operator in whatever form we decide to test / is necessary. 

To illustrate how this inherent flexibility can be used, we will now infer the relevant terms to include in the collision operator by performing a Pareto-curve study over the more general operator form described in Section~\ref{methods:K}. We will be varying the kernel size $k\times k$ of the general (non-local) operator while measuring the corresponding average test rollout error. For simplicity we will be using a uniform thermal distribution, and always consider a discrete version of the operator, but this methodology can be used for other scenarios and will be explored elsewhere.

\begin{figure}
    \centering
    \includegraphics[width=0.4\linewidth]{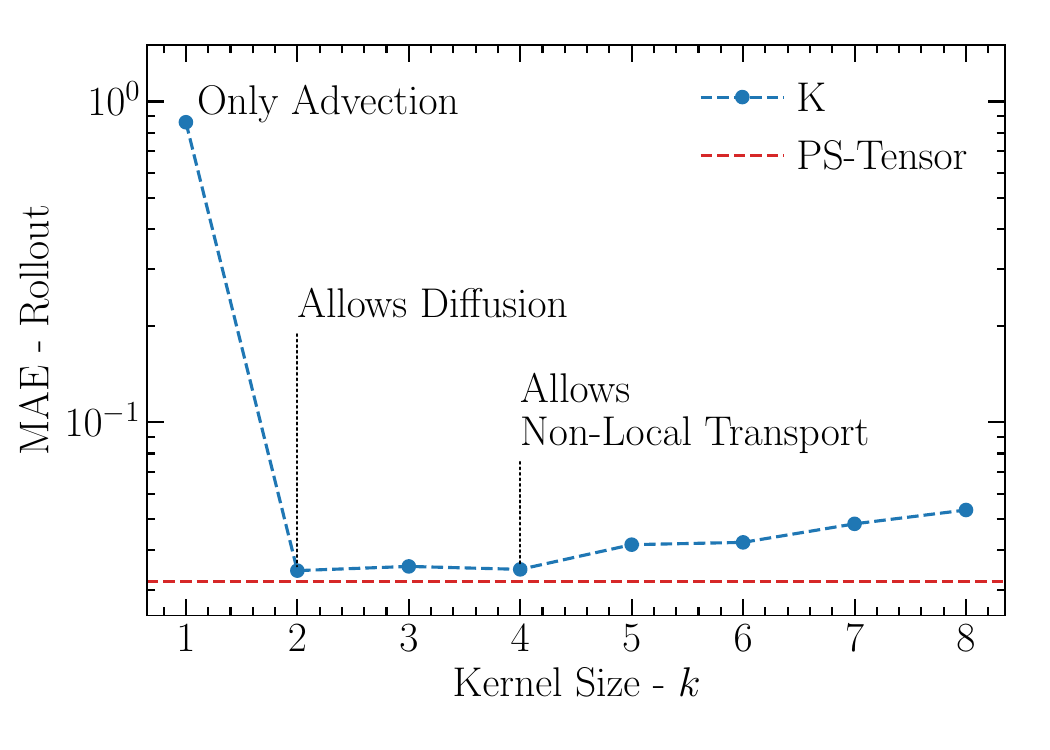}
    \caption{Pareto-curve on the impact of operator kernel size $k$ on rollout error using the corresponding integro-differential operator (K). For the simulation under study, the electron velocity distribution function is initialized using a thermal distribution with radius $v_{th} = 0.01c$ which does not vary over time. Numerical simulation parameters (which affect collisionality) correspond to a number of particles per cell $N_{ppc}=1$, linear particle shape function, and grid resolution $\Delta_x = 0.01 c/\omega_p$ (more details in~\ref{app:pic_simulation_parameters}). The baseline error for a pure advection-diffusion model trained on the simulation (PS-Tensor) is shown in red. The optimal value is $k=2$, which corresponds to an advection-diffusion operator. This is the expected result for collisional dynamics dominated by small-angle scattering collisions, as is the case for finite-size particle systems under the studied numerical parameters~\cite{langdon1970nonphysical, langdon1970theory, carvalho2026learning}. Further increasing $k$ leads to virtually no improvements and starts degrading the performance due to overfitting/training instabilities.}
    \label{fig:K-test_l1_avg}
\end{figure}

The Pareto-curve is shown in Figure~\ref{fig:K-test_l1_avg}. 
When considering $k=1$, the operator can only represent an advection term, and therefore can not accurately model the dynamics. For $k=2$, the operator can now include a diffusion term, thus significantly decreasing the rollout error. Further increasing the value of $k$ does not results in performance improvements, and in fact leads to worse performance. This confirms that, as expected, for the problem under study (non-relativistic thermal plasma composed of finite-size particles) an advection-diffusion operator is the most accurate description of collisional dynamics for numerical parameters under which small-angle scattering events dominate~\cite{langdon1970nonphysical, langdon1970theory, carvalho2026learning}. We can further test this point by training an equivalent PS-Tensor advection-diffusion model and observe that it achieves equivalent performance (red line) with respect to the kernel operator trained with $k=2$. Furthermore, we provide comparisons between the test phase space rollouts produced by the optimal kernel operator against the PIC simulation data in \ref{app:K} and find them to be in excellent agreement.

\begin{figure}
    \centering
    \includegraphics[width=0.7\linewidth]{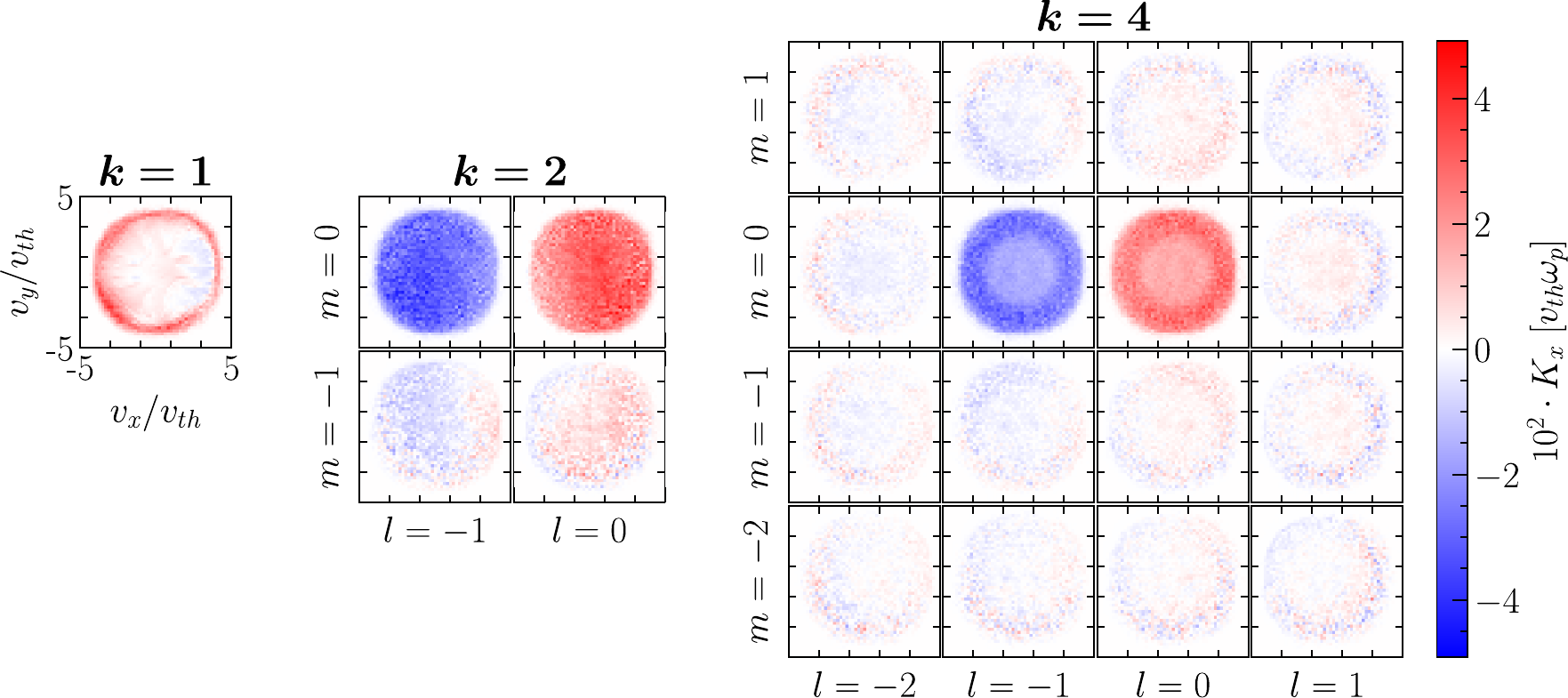}
    \caption{Discrete 4-Dimensional kernel operators recovered for different kernel sizes $k$ (correspond to the operators used in Figure~\ref{fig:K-test_l1_avg}). Only $K_x$ is shown since $K_x(v_x, v_y, l,m) = K_y(v_y, v_x, m, l)$ by construction. It is clear that for $k > 1$ there is a derivative term along $v_x$ that consistently dominates. This corresponds to a diffusion term, which is expected to be relevant for the dynamics of interest. The operator with $k=1$ can not compute this derivative and therefore fails to reproduce the phase space dynamics. Since no second derivative term appears for $k=4$, we can also infer that non-local transport is not relevant.}
    \label{fig:K_kernel}
\end{figure}

In Figure~\ref{fig:K_kernel} we showcase the learned operator form for $k=1,2,4$.
It is clear the appearance of a diffusion term in the form of a first derivative from $k=2$ onwards (note that this corresponds exactly to a diffusion term only if the extra gradient in the operator in eq.~\eqref{eq:K_operator} is computed with a forward difference method, with more details regarding the impact of the finite-difference scheme used in~\ref{app:K}). The (extra) unnecessary degrees of freedom are populated with small fluctuations which are the result of overfitting the training data and can actually lead to training instabilities for larger $k$ values. Although not explored in this work, it is possible to also regularize the solution to minimize the number of non-zero entries, which should produce a more accurate representation of the operator.

\section{Discussion}
The techniques proposed here can be applicable to both simulation~\cite{wong2025energy}, experimental \cite{bergeson2025experimental}, and observational~\cite{camporeale2022data} data. 
We expect that in the future, data-driven general form collisional operators will be integrated into existing codebases as an alternative to theoretical estimates. In addition, the ability to accurately infer time-dependent operators will be relevant for post-hoc analysis and reduced model development of complex stochastic plasmas processes, e.g. non-thermal particle acceleration in relativistic magnetized turbulence~\cite{alves2018efficient,wong2020first, wong2025energy}. These approaches offer new tools to test and even stimulate the development of new theoretical models of stochastic processes in plasma, from effective advection-diffusion operators for complex wave-particle interactions, to developing collision models in magnetized or strongly coupled relativistic plasmas.

It is important to note that the feasibility of the proposed approach depends on the diversity and quality of the available phase space diagnostics. For simulations, it is generally straightforward to produce high-quality phase space data for multiple subpopulations. However, a similar strategy might not be feasible for experimental or observational scenarios, making it harder to estimate unique operators for these cases. The impact of the number of subpopulations used to recover the operators is problem dependent and can not be determined \textit{a priori}. Nonetheless, it can be quantified, e.g., by following similar analysis to that performed in~\ref{app:discussion_non_uniqueness} where a convergence study is used to determine the minimum number of sub-populations required to obtain optimal performance, and a statistical ensemble of models is trained to estimate the uncertainty of the estimated coefficients.

Independently of the data source, the phase space diagnostic cadence is expected to resolve the expected (collisional) time-scales of interest. Given the reduced storage requirements necessary to store phase space snapshots compared to particle tracks, this is not expected to be a bottleneck. Furthermore, downsampling strategies (in time and/or space) can be later applied during post-hoc analysis to remove unnecessary information and reduce the computational burden of the inverse problem without affecting the results (or having to re-run simulations).

Data noise levels (due to stochastic measurement noise or lack of phase space statistics) are also expected to affect the applicability of the current approach since they affect the estimation of finite-difference derivatives and increase the ill-posedness of the problem. As demonstrated in~\cite{carvalho2026learning}, increasing the training rollout lengths and imposing physical symmetries are robust ways to address noisy phase space statistics. Future works could focus on proposing strategies to mitigate purely stochastic effects and devise optimized (general or possibly problem-dependant) operator forms that further mitigate these issues.

\section{Conclusions}

In this work we have extended the approach proposed in~\cite{carvalho2026learning} to be able to infer time-varying advection-diffusion operators from phase space data produced by self-consistent kinetic simulations. This is an important contribution since collisional and stochastic wave-particle dynamics in plasmas far from equilibrium are temporally evolving stochastic processes and, therefore, an operator which aims to capture their statistical effect (without information about the underlying background distribution and wave spectra) should also be time-dependent. We have also introduced a more general integral formulation of the operator, which can be used to infer meaningful functional contributions without explicitly implementing the corresponding terms.

Both approaches proposed in this paper are shown to correctly capture the operator that underlies the phase space dynamics. Importantly, we have also shown that for cases where collisional time-scales are similar to those of other collective processes (in this case plasma oscillations) and the advection-diffusion operator is evolving due to changes in the background distribution function, the proposed phase space method coupled with a differentiable simulator is capable of learning an accurate operator while estimates from particle tracks fail. We have further shown how a Pareto-curve analysis with the generalized operator form allows to identify an advection-diffusion model as the correct description for the finite-size particle plasma under study. These provide a powerful set of techniques to explore collisional effects and their relevance in different physical contexts.

The methodology proposed in this work establishes a foundation for exploring the complex interplay between collisional and stochastic wave-particle dynamics in plasmas using differentiable simulators. By further extending this framework, e.g., by including the effect of external fields and more general integral operator formulations that also take into account the wave spectra and background distribution function, we conjecture that new insights can be obtained into non-thermal particle acceleration and electromagnetically dominated collisional dynamics in both laboratory and astrophysical plasmas.

\ack{
The authors would like to thank S. Degen, G. Guttormsen, P. Bilbao, V. Decyk, and W. Mori for valuable discussions. The authors acknowledge the OSIRIS Consortium, consisting of UCLA, University of Michigan, and IST (Portugal) for the use of the
OSIRIS 4.0 framework.
}

\funding{
Simulations and machine learning workloads were performed in Deucalion (Portugal) within FCT Masers in Astrophysical Plasmas (MAPs) I.P project 2024.11062.CPCA.A3, FCT Machine-learned closures for plasma simulations I.P project 2024.12682.CPCA.A1, and EuroHPC proposal No. EHPC-DEV-2025D02-069.
This work was supported by FCT (Portugal) Grants No. 2022.13261.BD, No. 2022.02230.PTDC (X-MASER), and No. UIDB/FIS/50010/2020-PESTB 2020-23. DC research visit to the UCLA was sponsored by a Fulbright Grant for Research with the support of FCT and by the Mani L. Bhaumik Institute for Theoretical Physics at UCLA.
}


\data{The data that support the findings of this study are available upon reasonable request from the authors at the following URL/DOI: \href{https://doi.org/10.5281/zenodo.18863863‌}{https://doi.org/10.5281/zenodo.18863863‌}. The software developed is available at: ‌\href{https://github.com/diogodcarvalho/ml-pic-collision-operators}{https://github.com/diogodcarvalho/ml-pic-collision-operators}.
}


\clearpage

\appendix
\renewcommand{\thesection}{Appendix~\Alph{section}}
\clearpage

\section{Training procedure from phase space evolution}
\label{app:training_procedure}

All models trained using the differentiable simulator approach have access to the phase space dynamics of a set of N-subpopulations of particles ($N_{train}^{dists}$). The exact set of subpopulations used slightly varies depending on the Section/Appendix of the paper. Detailed information is provided in~\ref{app:subpopulation_sampling}. 

For all models shown in the paper we use all time-steps of all training subpopulations for training. Initial hyperparameter tuning was performed for the PS-NN operators in Section~\ref{sec:results_time_dependent_operator} by splitting the data into 90\% for training and 10\% for validation. However, after reasonably good performing values were found we stopped using this split. The optimized values correspond to MLPs with 3 layers, hidden dimension of 32, and LeakyReLU activation functions for all hidden layers.

To train the models, we use a curriculum learning approach as proposed in~\cite{carvalho2026learning}, where the temporal unrolling length $N_u$ is consistently increased. The different curriculum stages, their duration (number of epochs), and associated learning rates ($\alpha$) are summarized in Table~\ref{tab:training_curriculum}.

The loss function computed at each stage is the mean absolute error over the temporal unrolled phase space dynamics:
\begin{equation}
    \mathcal{L}_{N_u} = \frac{1}{N_uN_v^2}\sum_{u=1}^{N_u}\sum_{v_x,v_y \in \mathcal{V}}\left|\hat{f}^{(t+u)}(v_x, v_y) - f^{(t+u)}(v_x,v_y)\right|
\end{equation}
averaged over the batch size, where $N_v^2$ corresponds to the total number of bins of phase space $\mathcal{V}$.

For all cases, we use the full training set per gradient update (i.e., batch size equals the training set size) and the Adam optimizer~\cite{kingma2014adam}. The final stored model weights correspond to the ones that achieved lower training loss at the last curriculum stage. Training times vary depending on the usage of a discrete or NN approach, the enforced symmetries, and the length of the simulation. Typical values are between 5-60 minutes on a single Nvidia A100 without using PyTorch~\cite{ansel2024pytorch} just-in-time compilation capabilities.

\begin{table}
\begin{center}
\begin{tabular}{c|ccc|ccc|ccc}
\hline
 & \multicolumn{3}{c|}{PS-Tensor (Section~\ref{sec:results_time_dependent_operator})} & \multicolumn{3}{c|}{PS-NN (Section~\ref{sec:results_time_dependent_operator})} & \multicolumn{3}{c}{K (Section~\ref{sec:results_K})} \\ \hline
Stage & $N_u$ & $\alpha$ & Epochs & $N_u$ & $\alpha$ & Epochs & $N_u$ & $\alpha$ & Epochs  \\
\hline
1 & 1  & $10^{-2}$ & 1000 & 1  & $10^{-3}$ &  500 & 1  & $10^{-2}$ & 500 \\
2 & 2  & $10^{-2}$ & 200  & 1  & $10^{-4}$ &  500 & 1  & $10^{-3}$ & 500 \\
3 & 5  & $10^{-2}$ & 200  & 2  & $10^{-4}$ &  200 & 2  & $10^{-3}$ & 200 \\
4 & 10 & $10^{-2}$ & 500  & 5  & $10^{-4}$ &  200 & 5  & $10^{-3}$ & 200 \\
5 &    &           &      & 10 & $10^{-4}$ &  500 & 10 & $10^{-3}$ & 500 \\
\hline
\end{tabular}
\caption{Curriculum training stages for the different models used throughout this work. The temporal unrolling length $N_u$ is increased progressively, the learning rate $\alpha$ and the number of epochs are adapted accordingly to ensure convergence and stability.}
\label{tab:training_curriculum}
\end{center}
\end{table}

\section{PIC Simulation Parameters}
\label{app:pic_simulation_parameters}

The detailed list of the simulation parameters used in this work is given in Table~\ref{tab:pic_simulation_parameters}. Both Sim-1 and Sim-2 consider an initial isotropic waterbag distribution and are relevant for the results in Section~\ref{sec:results_time_dependent_operator} and~\ref{app:time_dependent_operator}. Sim-3 considers a thermal velocity distribution and is relevant for Section~\ref{sec:results_K} and~\ref{app:K}. Across all simulations the time-step was chosen as $c\Delta t = 0.98\Delta_x/\sqrt{2}$ to ensure that the CFL stability condition was fulfilled~\cite{courant1928partiellen}.

\begin{table}
\caption{PIC simulation parameters used. Parameters correspond to: $v_w$ - initial waterbag distribution radius; $v_{th}$ - thermal velocity, $N_{ppc}$ - number of particles per cell; $m$ - shape function order; $x_{max}$ - box size, equal for both directions; $\Delta_x$ - grid resolution, equal for both directions; $t_{max}$ - maximum simulation time; $\Delta t_{dump}$ - particle and phase space diagnostic period. All velocities are normalized to the speed of light ($c$) and time intervals to the inverse of the plasma frequency (1/$\omega_p$), resulting in distances in units of $c/\omega_p$.}
\centering
\begin{tabular}{l c c c c c c c c c}
\hline
Simulation Identifier & $v_w$ & $v_{th}$ & $N_{ppc}$ & $m$ & $x_{max}$ & $\Delta_x$ & $t_{max}$ & $\Delta t_{dump}$ \\
\hline
Sim-1 (Section~\ref{sec:results_time_dependent_operator}, \ref{app:time_dependent_operator}) & 0.02 & -    & 4  & 1 &  100 & 0.01 & 100 & 0.53 \\
Sim-2 (\ref{app:time_dependent_operator}) & 0.02 & -    & 25 & 2 &  50  & 0.01 & 730 & 3.65 \\
Sim-3 (Section~\ref{sec:results_K}, \ref{app:K}) & -    & 0.01 & 1  & 1  & 100 & 0.01 & 100 & 1.00 \\
\hline
\end{tabular}
\label{tab:pic_simulation_parameters}
\end{table}

\section{Subpopulation Sampling}
\label{app:subpopulation_sampling}

To generate the phase space evolution of different subpopulations we follow the procedure described in~\cite{carvalho2026learning}. We sample particles at $t=0$ according to a desired distribution function and store the phase space evolution of their corresponding density distribution. For all simulations in Table~\ref{tab:pic_simulation_parameters} we discretize the phase space $\mathcal{V} = [-0.05,0.05]c\times[-0.05,0.05]c$ over a uniform grid of size $N_v\times N_v = 51\times51$ bins. 

For Sim-3 (Section~\ref{sec:results_K}) we use the exact same train/test distributions as described in~\cite{carvalho2026learning}. These correspond to 9 train subpopulations (sampled from normal distributions centred at different positions) and 19 test subpopulations (distinct normal distributions, rings, quadrants, etc). 

For Sim-1 and Sim-2 (Section~\ref{sec:results_time_dependent_operator}) we use a different (smaller) set. These are:
\begin{itemize}
    \item uniform$_{-5,5,-5,5}$ : Uniform distribution covering $V: [-5,5] \times [-5,5]$
    \item uniform$_{-5,0,-5,0}$ : Uniform distribution covering $V: [-5,0] \times [-5,0]$
    \item normal$_{0,0}$: Normal distribution centered at $\boldsymbol{\mu} = (0,0)$ and covariance $\Sigma_{ii} = 0.05$ and $\Sigma_{ij} = 0$
    \item normal$_{-1,-1}$: Equivalent with center at $\boldsymbol{\mu} = (-1,-1)$
    \item normal$_{-1,0}$: Equivalent with center at $\boldsymbol{\mu} = (-1,0)$ 
    \item normal$_{-2,-2}$: Equivalent with center at $\boldsymbol{\mu} = (-2,-2)$ 
    \item normal$_{-2,0}$: Equivalent with center at $\boldsymbol{\mu} = (-2,0)$ 
    \item normal\_rot$_{0}$ : Normal distributions centered at $\boldsymbol{\mu} = (0,0)$ and covariance matrix with entries $\Sigma_{xx}= 1$, $\Sigma_{yy}=0.05$, and $\Sigma_{xy} = 0$
    \item normal\_rot$_{45}$ : Equivalent but rotated by $\theta = 45^{\circ}$
    \item ring$_{0.5}$: Ring centered at radius $\mathrm{v} = 0.5$ with standard deviation $\sigma_\mathrm{v} = 0.2$
    \item ring$_{1.0}$: Equivalent centered at radius $\mathrm{v} = 1.0$ 
    \item ring$_{1.5}$: Equivalent centered at radius $\mathrm{v} = 1.5$   
\end{itemize}
where all velocities are normalized to $v=0.01c = v_w/2$. For all distributions we sample a total of $3\times10^5$ particles. All subpopulations are used for test purposes but the training set varies throughout the paper. Refer to Table~\ref{tab:waterbag_training_dists} for more details.

\begin{table}
\caption{Subpopulations sampled for studies performed using an initial waterbag velocity distribution. All subpopulations are used for testing purposes. Subpopulations used for training vary depending on the Section and Appendix. The PS-Tensor and PS-NN models shown in the main body of the paper all use $N_{train}^{dists}=7$.}
\centering
\resizebox{\textwidth}{!}{\begin{tabular}{l c c c c c c c}
\hline
Identifier              & $N_{dists}^{train}=1$ & $N_{dists}^{train}=2$  & $N_{dists}^{train}=3$ & $N_{dists}^{train}=5$ & $N_{dists}^{train}=7$ & $N_{dists}^{train}=9$ & $N_{dists}^{train}=12$ \\
\hline
uniform$_{-5,5,-5,5}$  & \checkmark & \checkmark & \checkmark & \checkmark & \checkmark & \checkmark & \checkmark\\
uniform$_{-5,0,-5,0}$ & & \checkmark & \checkmark & \checkmark & \checkmark & \checkmark & \checkmark\\
normal$_{0,0}$ & & & \checkmark & \checkmark & \checkmark & \checkmark & \checkmark\\
normal$_{-1,-1}$ & & & &\checkmark & \checkmark & \checkmark & \checkmark\\
normal$_{-1,0}$  & & & &\checkmark & \checkmark & \checkmark & \checkmark\\
normal$_{-2,-2}$ & & & & &\checkmark & \checkmark & \checkmark\\
normal$_{-2,0}$ & & & & &\checkmark & \checkmark & \checkmark\\
normal\_rot$_0$ & & & & & &\checkmark & \checkmark\\
normal\_rot$_{45}$ & & & & & &\checkmark & \checkmark\\
ring$_{0.5}$  & & & & & & & \checkmark\\
ring$_{1.0}$  & & & & & & &\checkmark\\
ring$_{1.5}$  & & & & & & &\checkmark\\
\hline
\end{tabular}}
\label{tab:waterbag_training_dists}
\end{table}

\section{Time-Dependent Advection-Diffusion Operator - Extra}
\label{app:time_dependent_operator}

\subsection{Estimation of coefficients from particles tracks}
\label{app:estimation_of_coefficients_from_tracks}

To estimate advection diffusion coefficients from particle tracks we follow a similar procedure as described in~\cite{carvalho2026learning} now modified to account for advection-diffusion coefficients which are changing over time. First we discretize the phase space into the same grid of $N_v\times N_v$ bins used for the phase space evolution of different subpopulations. Advection and diffusion coefficients for each bin are estimated by computing:
\begin{equation}
    A_i(t,\boldsymbol{v}_{bin}) = \frac{<\Delta v_i>_{\boldsymbol{v}(t) \in \boldsymbol{v}_{bin}}}{\Delta t}
\label{eq:a_tracks}
\end{equation}
\begin{equation}
    D_{ij}(t,\boldsymbol{v}_{bin}) = \frac{<\Delta v_i\Delta v_j>_{\boldsymbol{v}(t) \in \boldsymbol{v}_{bin}}}{\Delta t}
\label{eq:d_tracks}
\end{equation}
where the under-scripts $i,j$ represent the corresponding axis, $t$ the time-step at which the coefficients are being computed, $\boldsymbol{v}_{bin}$ all velocities contained inside the phase space bin, $\Delta v_i =v_i(t+\Delta t) - v_i(t)$ an individual particle velocity change over the time interval $[t,~t+\Delta t]$, and $<.>_{\boldsymbol{v}\in \boldsymbol{v}_{bin}}$ an average over all particles in the phase space bin at timestep $t$. Throughout this work, we usually plot the projections of $A_i, D_{ij}$ onto their parallel and perpendicular components with respect to the bin average particle velocity for ease of comparison against operators learned using the differentiable simulator (e.g. in Figure~\ref{fig:AD-sim1-comparisons}).

As demonstrated in~\cite{carvalho2026learning} the choice of $\Delta t$ can significantly affect the estimated values of $A_i,D_{ij}$ and therefore influence the accuracy of the retrieved operator. Furthermore, unlike in~\cite{carvalho2026learning}, the background distribution is now evolving over time, which is expected to further limit the possibility of performing averages over long periods of times since $A_i$ and $D_{ij}$ are changing.

To provide the best possible estimate of the coefficients from particle tracks we analyzed the impact of: 1) the time-step over which statistics are computed, i.e. $\Delta t=N_{dump}t_{dump}$ where $t_{dump}$ is the diagnostic dumping period; 2) using overlapping vs non-overlapping time-intervals. When using overlapping time-steps, $A_i$ and $D_{ij}$ are defined at $t \in\{0,1,...,N_{max}-N_{dump}\}$ while for non-overlapping they are defined at $t \in\{0, N_{dump}, ..., \lfloor N_{max}/N_{dump}\rfloor - N_{dump}\}$ where $\lfloor .\rfloor$ represents the floor function. During test rollouts, the values for time-steps over which the coefficients are not defined are computed via linear interpolation from neighboring time-steps.

\begin{figure}
    \centering
    \includegraphics[width=0.4\linewidth]{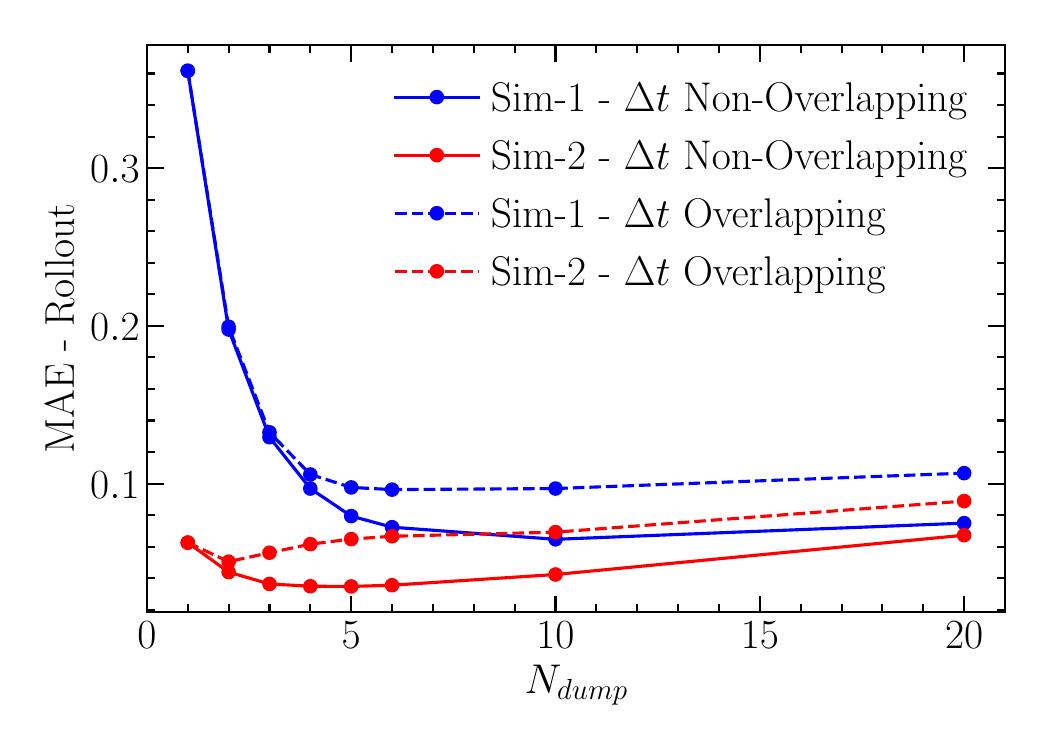}
    \caption{Rollout error comparison when computing advection-diffusion coefficients from particle tracks using either non-overlapping or overlapping time-intervals of size $\Delta t=N_{dump}\Delta t_{dump}$. Using non-overlapping time-steps produces the best results and is the strategy used for comparisons throghout the paper.}
    \label{fig:waterbbag_tracks_overlapping_vs_nonoverlapping}
\end{figure}

\begin{figure}
    \centering
    \includegraphics[width=0.4\linewidth]{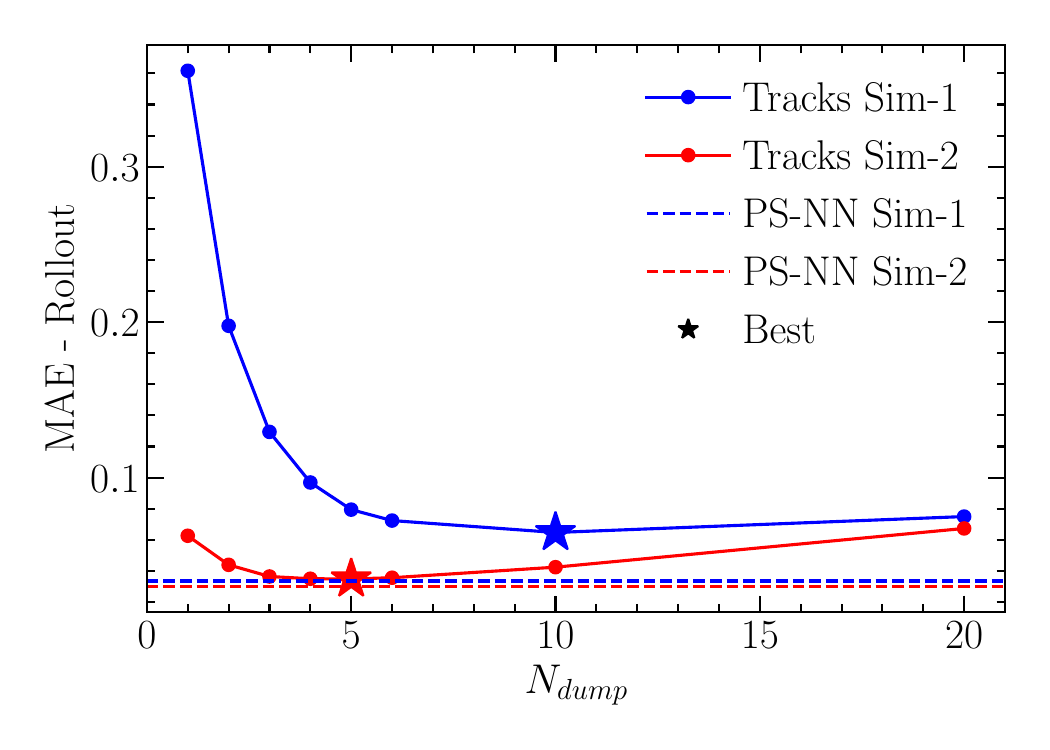}
    \caption{Rollout error for operators retrieved from particle tracks in function of $N_{dump}$, the number of diagnostic time-steps over which $<\Delta v_i>$ and $<\Delta v_i \Delta v_j>$ are calculated. Typical MAE-Rollout errors for PS-NN method for each simulation are provided for comparison. For Sim-1 the coefficients measured from particle tracks do not capture the phase space dynamics accurately as demonstrated by the higher error. This happens because collisional dynamics occur on similar time-scales as plasma oscillations which biases the measurements. For Sim-2 there is a larger scale separation so the problem does not manifest. The phase space method based on the differentiable simulator approach successfully recovers the correct operator in both cases. Operators marked with $\star$ are used for comparisons throughout the manuscript.}
    \label{fig:waterbag_tracks_ndump_error}
\end{figure}

In Figure~\ref{fig:waterbbag_tracks_overlapping_vs_nonoverlapping} we demonstrate the rollout error obtained in function of both $N_{dump}$ and of computing statistics over overlapping $\textit{versus}$ non-overlapping intervals. 
It is clear that using non-overlapping time-steps provides operators which achieve smaller errors, while the optimal value of $N_{dump}$ is simulation dependent. It is important to highlight that the minimum error achieved for Sim-1 (at $N_{dump}=10$, which corresponds to $\Delta t\omega_p\approx5$) is still large, which indicates that the estimated operator is nonetheless incorrect. To further highlight this point, we show in Figure~\ref{fig:waterbag_tracks_ndump_error} a comparison of non-overlapping scenario against the errors obtained by operators retrieved using the differentiable simulator.
While for Sim-2 we observe similar errors between the best performing value of $N_{dump}$ (corresponds to $\Delta t\omega_p \approx 18)$  and the PS-NN case, for Sim-1 there exists a large gap in performance. 

Similarly to our justifications in~\cite{carvalho2026learning} we believe this is due to the plasma oscillation scales being close to the collisional time-scales for Sim-1. This is even more problematic for the time-varying coefficients because we can not perform statistics over larger time intervals since the operator is evolving over time. We believe these results further highlight the benefits of the data-driven approach proposed in this work which seems capable of avoiding this pitfall. 

\subsection{Extra rollout examples}
\label{app:ad_extra_rollouts}

\begin{figure}
    \centering
    \includegraphics[width=0.57\linewidth]{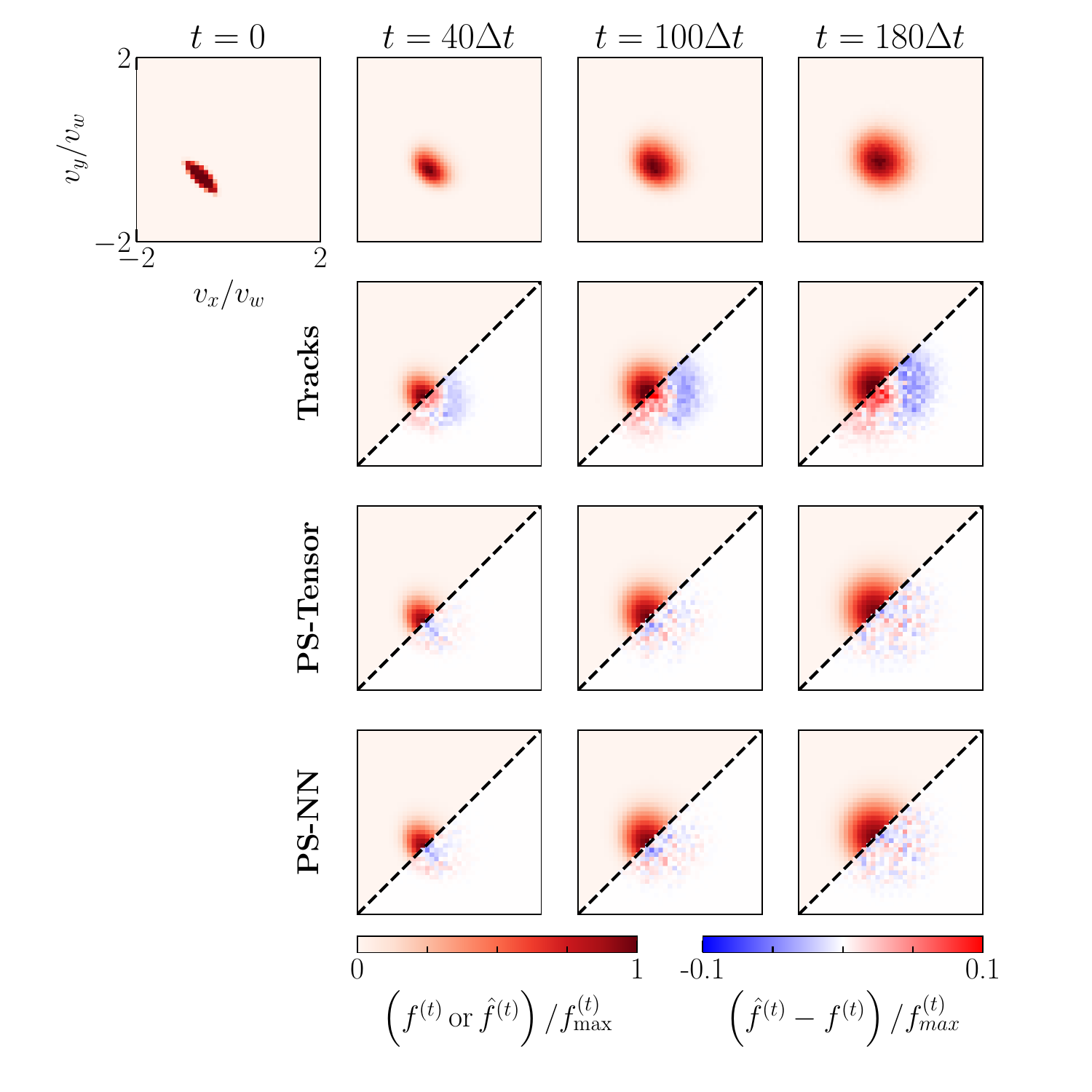}
    \includegraphics[width=0.57\linewidth]{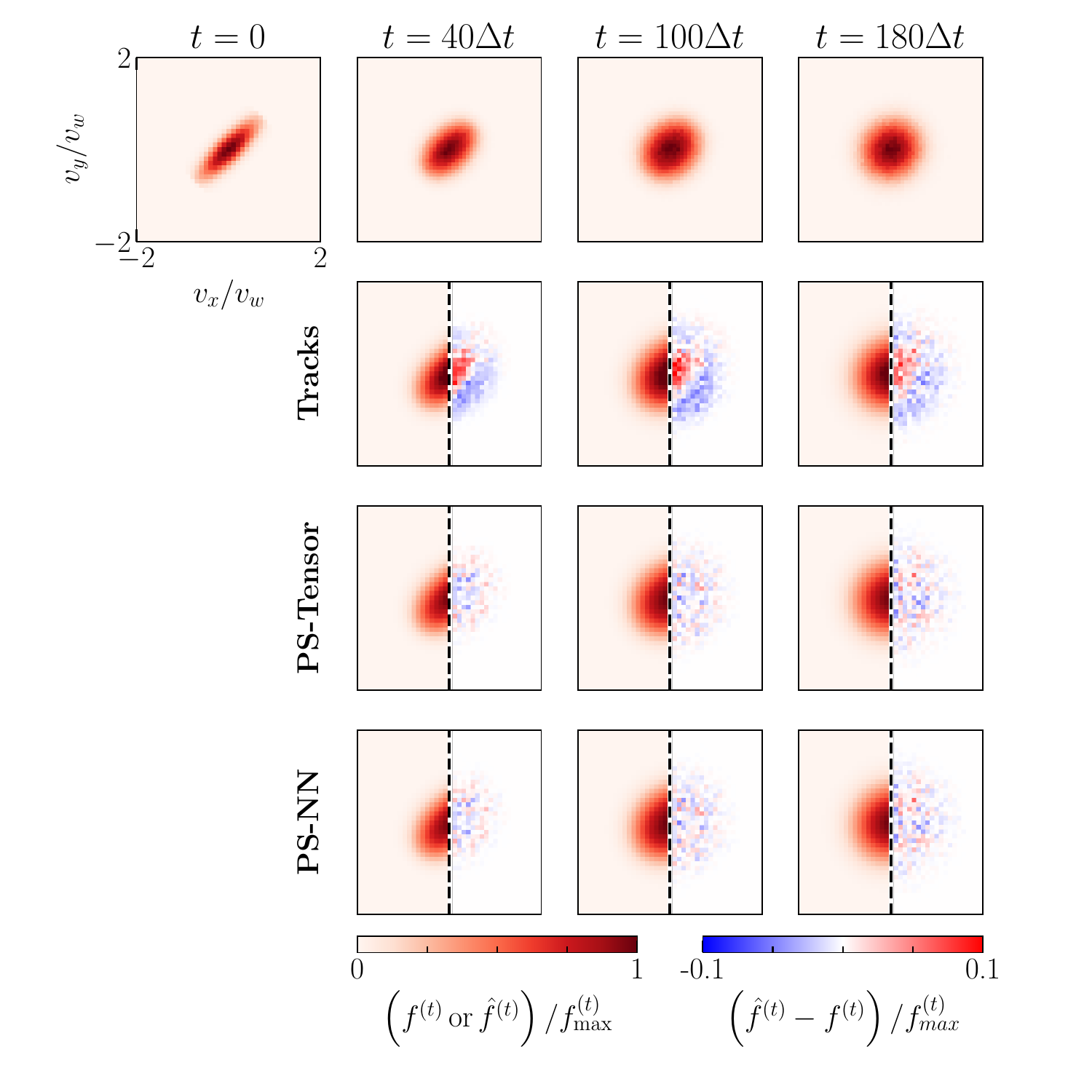}
    \caption{Additional comparisons between phase space evolution of different subpopulations for different operators retrieved in Section~\ref{sec:results_time_dependent_operator}. It is clear that the operator retrieved from particle tracks can not accurately reproduce the phase space dynamics. On the other hand, the phase space based methods accurately reproduce the long term dynamics.}
    \label{fig:AD-sim1-extra-examples-1}
\end{figure}

\begin{figure}
    \centering
    \includegraphics[width=0.57\linewidth]{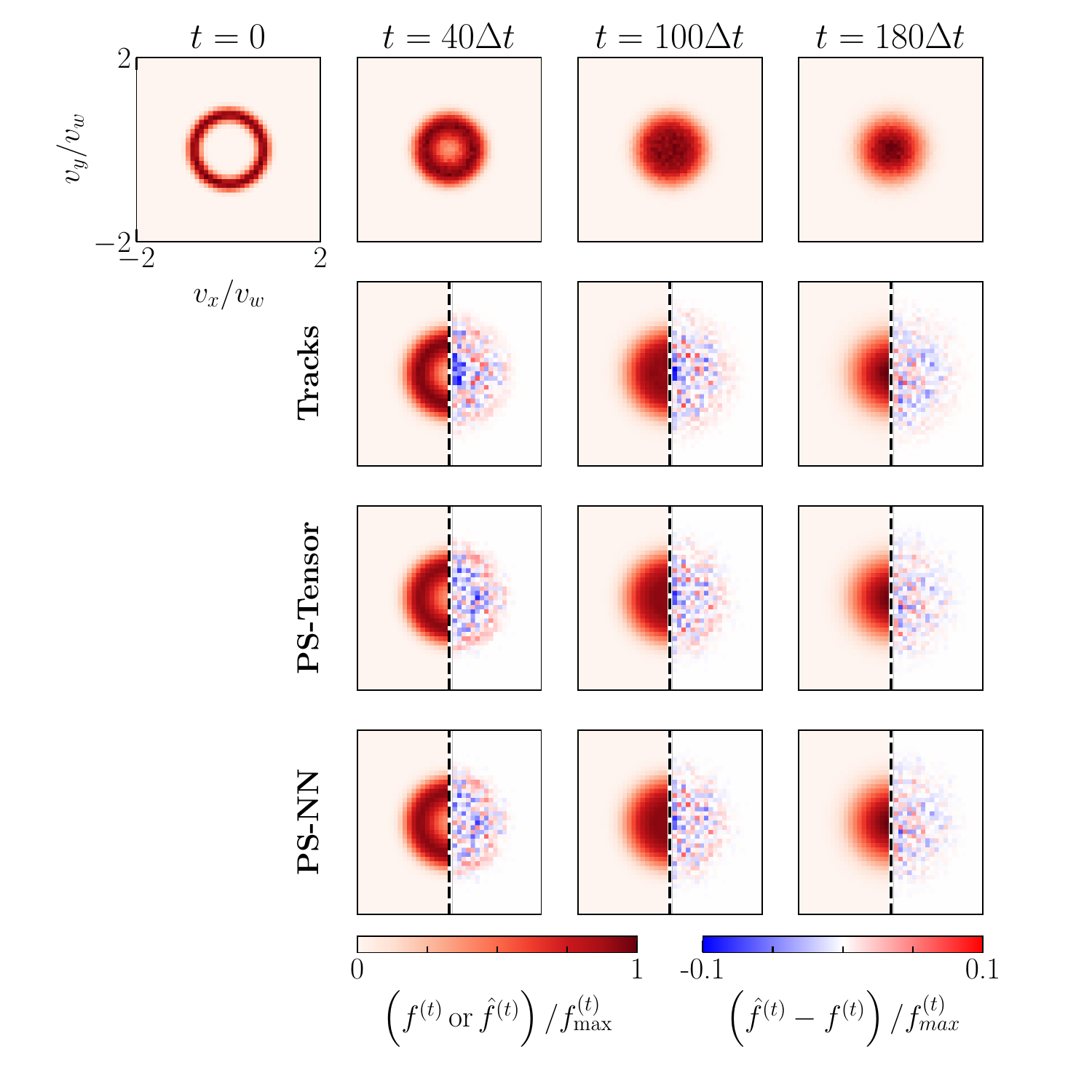}
    \includegraphics[width=0.57\linewidth]{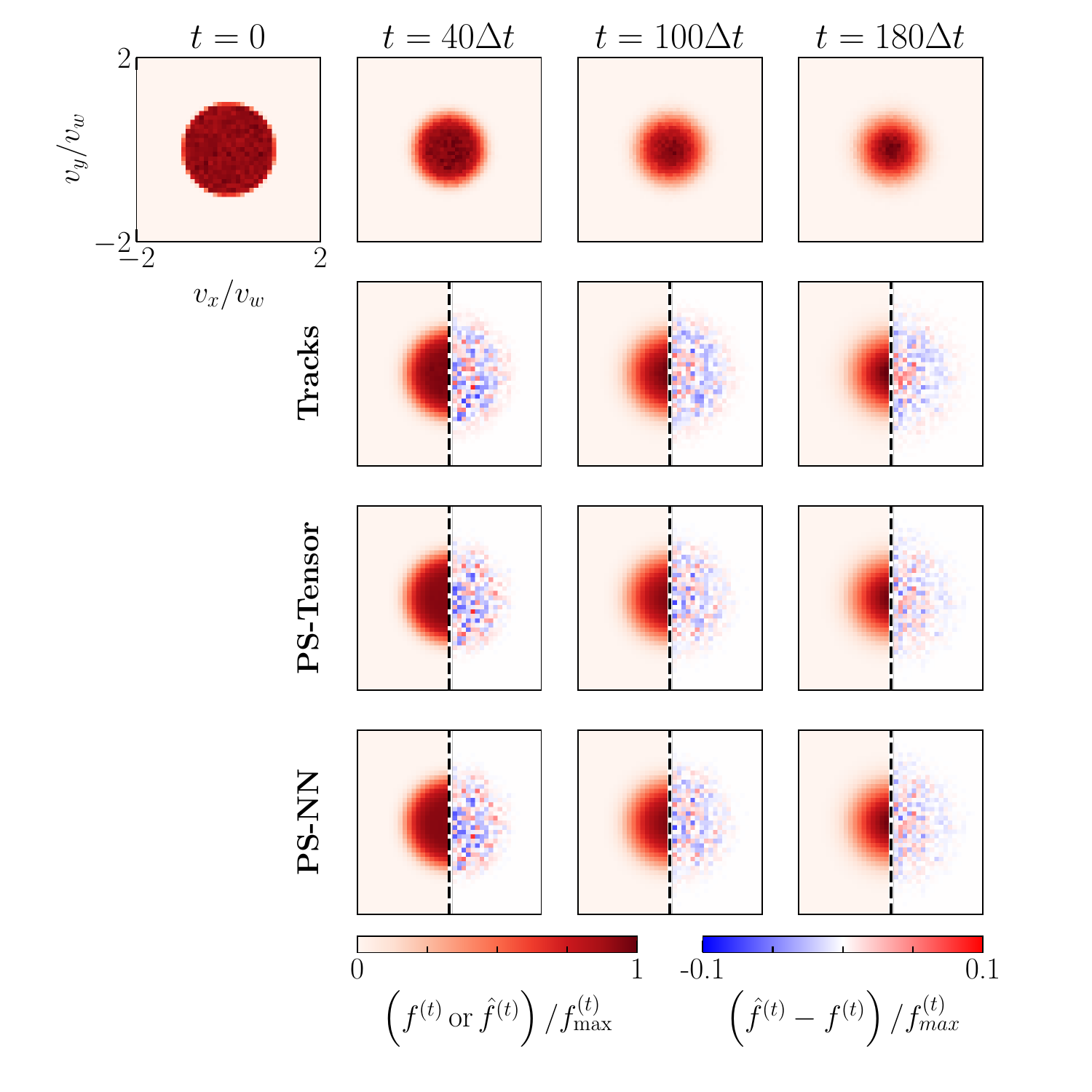}
    \caption{Additional comparisons between the phase space evolution of different subpopulations for the different operators retrieved in Section~\ref{sec:results_time_dependent_operator}. For these cases it is not as clear the worse performance of the Tracks operator since distributions are already isotropic and the discrepancy in the diffusion terms at low $t$ do not play a significant role (with the clear exception for the ring distribution at $t=40\Delta t$).}
    \label{fig:AD-sim1-extra-examples-2}
\end{figure}

In this Appendix we provide in Figures~\ref{fig:AD-sim1-extra-examples-1} and \ref{fig:AD-sim1-extra-examples-2} further examples of comparisons between the phase space dynamics of different subpopulations obtained from PIC \textit{versus} those simulated using different estimated operators.
These results are complementary to those shown in Section~\ref{sec:results_time_dependent_operator}. The new plots further highlight the errors produced by a wrong estimate of diffusion by the Tracks method, while showcasing the overall good agreement between the PS-Tensor and PS-NN results with respect to the original PIC data.

\subsection{Results for simulation with smaller collisionality}
\label{app:sim-2}

In this Appendix we reproduce the results from Section~\ref{sec:results_time_dependent_operator} now for a simulation with reduced collisionallity (by increasing the number of particles per cell and shape function order, see Table~\ref{tab:pic_simulation_parameters} for a comparison). 

\begin{figure}
    \centering
    \includegraphics[width=0.8\linewidth]{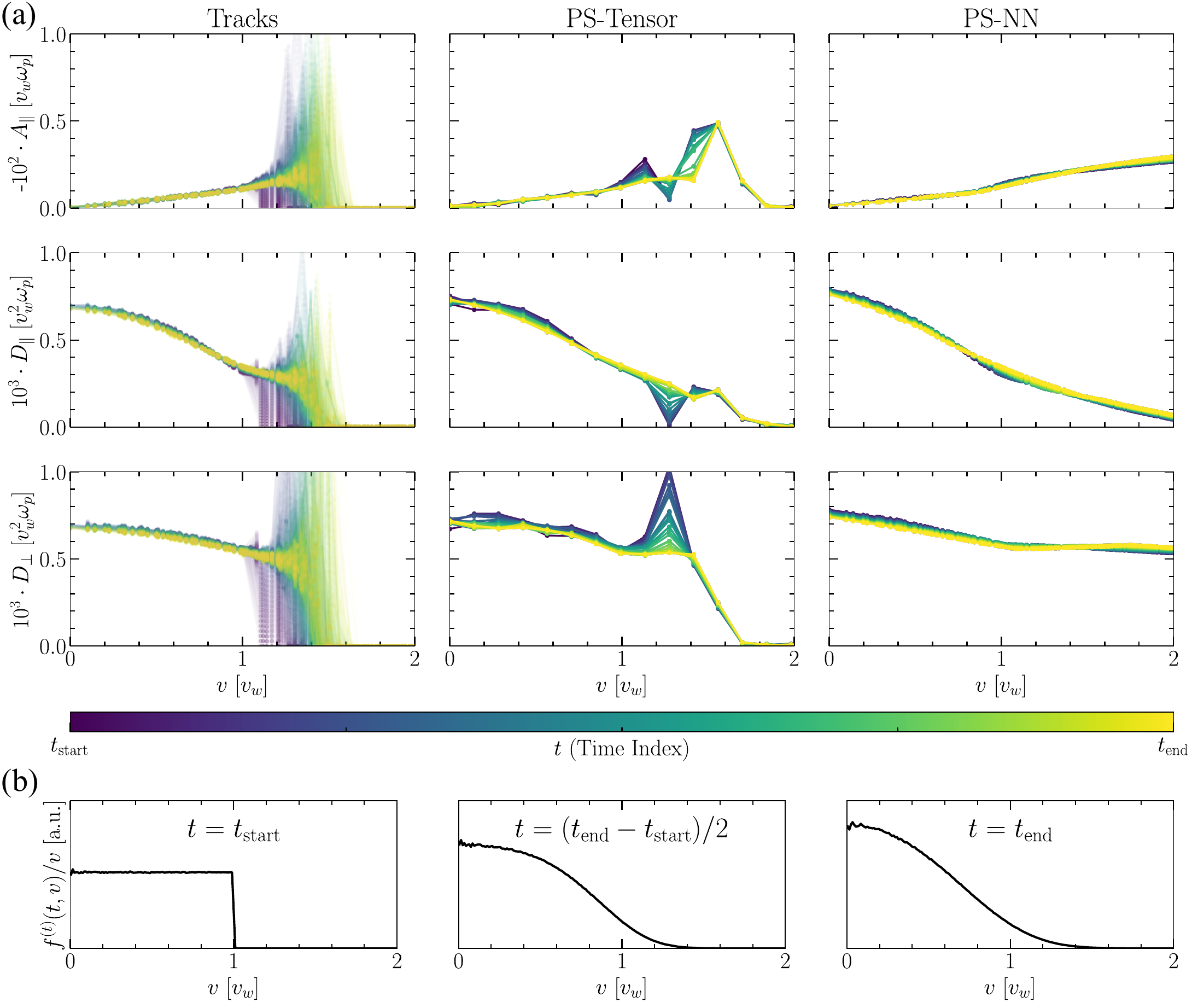}
    \caption{(a) Time-dependent advection-diffusion coefficients retrieved using  Tracks, and phase space based approaches with a discrete (PS-Tensor) and continuous (PS-NN) function approximators for Sim-2 ($N_{ppc}=25$, quadratic shape function $m=2$). Unlike for Sim-1 ($N_{ppc}=4$, linear shape function $m=1$; shown in Figure~\ref{fig:AD-sim1-comparisons}(a) in the main body of the paper) there is much better agreement between estimates from tracks and phase space models. (b) Snapshots of the velocity distribution function. It is clear that the background waterbag distribution evolves similarly over the full simulation, when compared to Sim-1 (shown in Figure~\ref{fig:AD-sim1-comparisons}(b) in the main body of the paper). Since the diagnostic frequencies for Sim-1 and Sim-2 is are matched to the same fixed fraction of the maximum simulation time (see Table~\ref{tab:pic_simulation_parameters}) the evolution of the background distribution function is similar across consecutive track/phase space diagnostic samples. This demonstrates that the incorrect estimates obtained for the advection-diffusion models from particle tracks for Sim-1 are not related to an unresolved fast change of the background distribution function, since the same issue is not observed for Sim-2. What varies between the two simulations is the scale separation between collisional dynamics and the plasma oscillation period, which we conjecture to be the cause of the problems observed for Sim-1.}
    \label{fig:AD-sim2-comparisons}
\end{figure}

\begin{figure}
    \centering
    \includegraphics[width=0.57\linewidth]{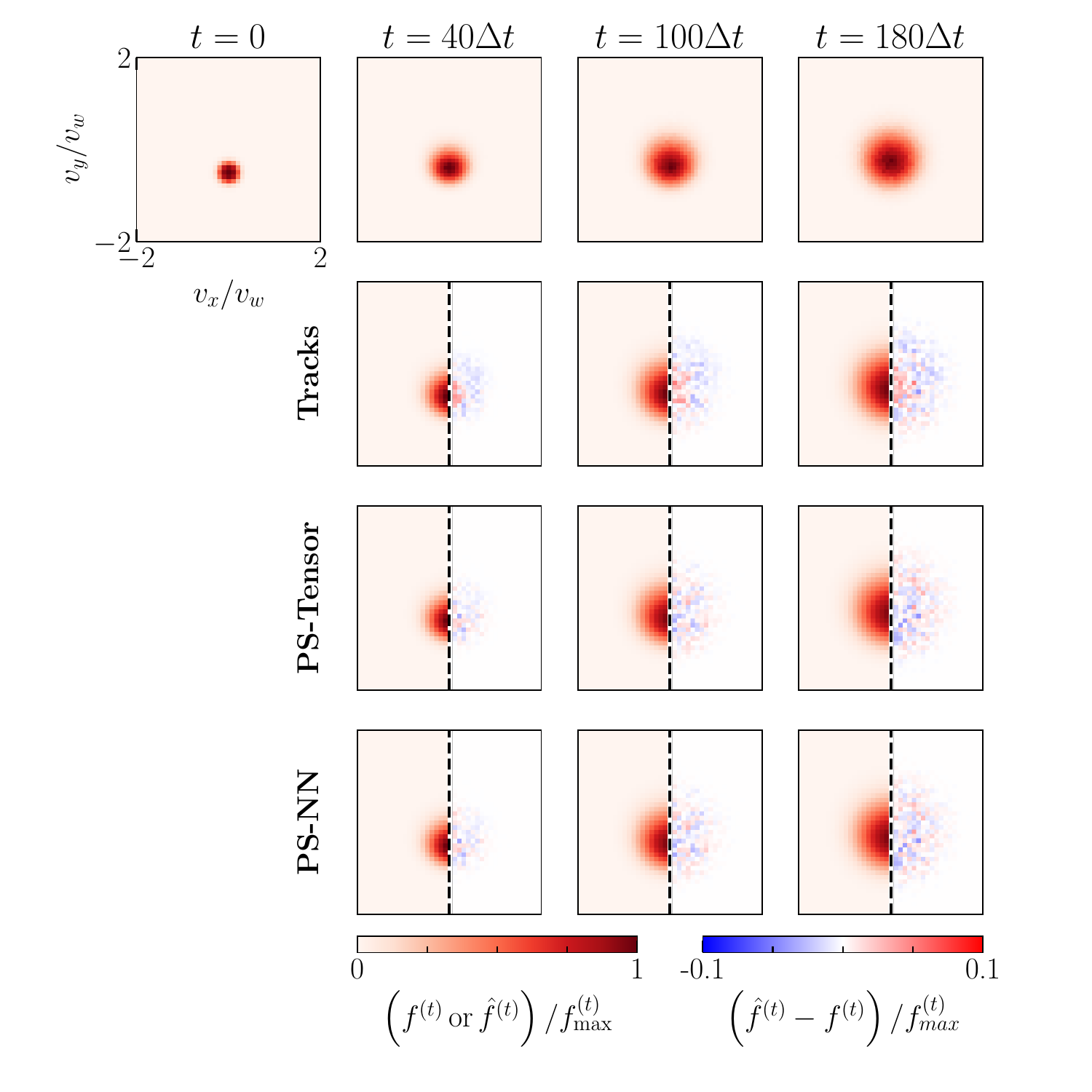}
    \caption{Example of the phase space evolution of a subpopulation for Sim-2 (cf. Table~\ref{tab:pic_simulation_parameters}). We include the results from the PIC simulation (top row) as well as the dynamics simulated using the learned operators (Tracks to PS-NN) shown in Figure~\ref{fig:AD-sim2-comparisons}. The error associated with the operator from Tracks has been reduced when compared to the results for Sim-1 in Figure~\ref{fig:AD-waterbag-ex_rolllout_dif_normal_-1_0} but there is still a smaller systematic error showcased by the consistent blue/red regions. The PS-Tensor and PS-NN models perform better, whith equivalent errors among them.}
    \label{fig:AD-waterbag-ex_rolllout_dif_normal_-1_0-sim2}
\end{figure}

The obtained advection-diffusion models from particle tracks and phase space methods are presented in Figure~\ref{fig:AD-sim2-comparisons}. It is clear that there is now a much better agreement between the operators estimated from particles tracks and those estimated from the phase space dynamics. We also observe a much smaller gap in terms of the average errors when predicting the long term dynamics (compare errors in Table~\ref{tab:sim-2-error} with Table~\ref{tab:sim-1-error}) and an equivalent subpopulation phase space rollout (compare Figure~\ref{fig:AD-waterbag-ex_rolllout_dif_normal_-1_0-sim2} with Figure~\ref{fig:AD-waterbag-ex_rolllout_dif_normal_-1_0}). 

These results highlight the benefits of learning operators using the proposed approach based on the phase space evolution of subpopulations instead of particle tracks, particularly for strongly coupled systems where collisional time-scales are closes to those of other collective processes.

\begin{table}
\centering
\caption{Average rollout error over multiple initial test subpopulations for operators retrieved for Sim-2 (operators in Figure~\ref{fig:AD-sim2-comparisons}). The error difference between operators retrieved from tracks versus phase space is significantly smaller than the one observed for Sim-1 (Table~\ref{tab:sim-1-error}) where operators learned from Tracks could not accurately reproduce the dynamics.}

\label{tab:sim-2-error}
\begin{tabular}{lccc}
\hline
 & Tracks & PS-Tensor & PS-NN \\
\hline
MAE-Rollout $\times 10^2$  & $3.48 \pm 0.42$ & $3.03 \pm 0.24$ & $3.00 \pm 0.26$ \\
\hline
\end{tabular}
\end{table}

\subsection{Discussion on the non-uniqueness of the operator}
\label{app:discussion_non_uniqueness}

Determining the advection-diffusion operator is an ill-posed problem since there is a family of solutions which can reproduce the same phase space dynamics. The proposed way to constrain the space of solutions is to provide the models with a varied set of distribution functions during train time~\cite{carvalho2026learning}. In this appendix we test this hypothesis. We evaluate if increasing the number of distribution functions used during training ($N^{train}_{dists}$) leads to: 1) improved generalization during test time; 2) convergence to a unique solution. 

For this purpose we do a scan on $N^{train}_{dists}$, and for each value train 8 equivalent NN models initialized with different random seeds. The training distribution functions used for each value of $N_{dists}^{train}$ are specified in Table~\ref{tab:waterbag_training_dists}. We highlight that the case of $N_{dists}^{train}=1$ corresponds to using an isotropic waterbag distribution that is equivalent to the full background distribution function (containing now only $\approx 7\%$ of the particles) and therefore should cover the full phase space where test dynamics will occur. Across all cases we are considering Sim-1 from Table~\ref{tab:pic_simulation_parameters}.

\begin{figure}
    \centering
    \includegraphics[width=0.4\linewidth]{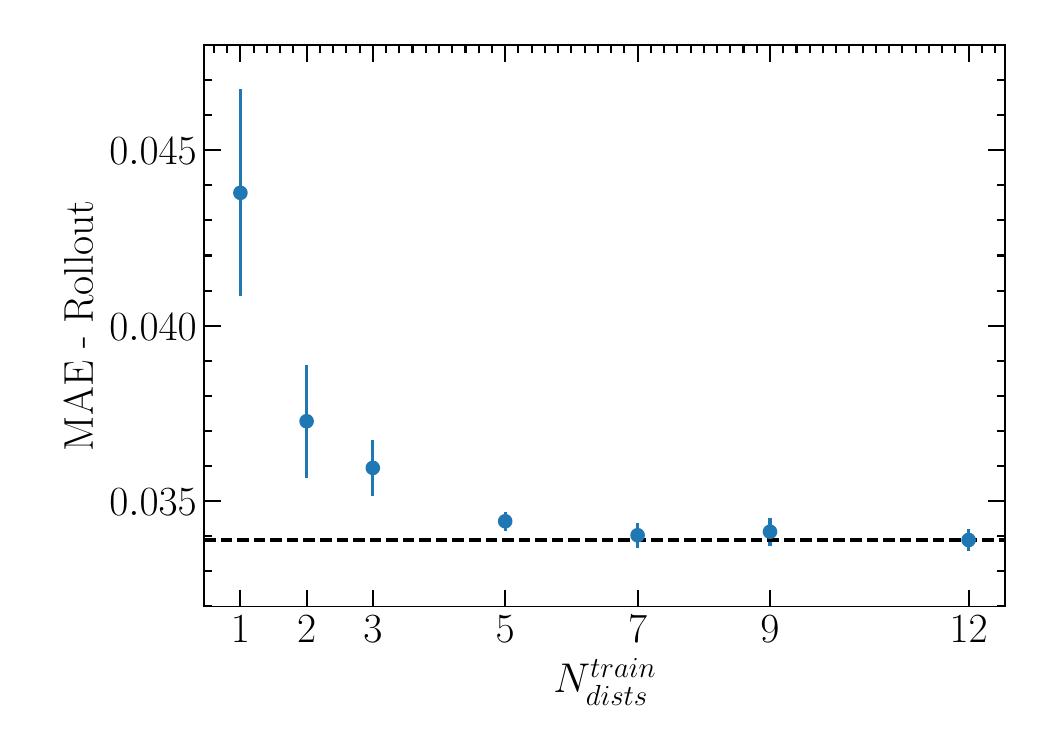}
    \caption{Test rollout error in function of the number of training distributions for PS-NN models for Sim-1 (cf. Table~\ref{tab:pic_simulation_parameters}). Markers represent average and standard deviations across 8 different initial random seeds. The horizontal dashed line represents the best average error achieved. Using $N_{dists}^{train} \geq 7$ does not produce significant improvements in performance.}
    \label{fig:n_train_dists-l1_avg-sim1-nn}
\end{figure}

In Figure~\ref{fig:n_train_dists-l1_avg-sim1-nn} we show the rollout error across all models trained. It is clear that increasing the number of training distributions leads to a more accurate estimate of the coefficients, with a plateau for $N_{dists}^{train} \geq 7$ (we observed also equivalent behaviour for PS-Tensor models). Furthermore, the standard deviation across models also decreases, indicating that models generalize similarly at test time.

\begin{figure}
    \centering
    \includegraphics[width=0.8\linewidth]{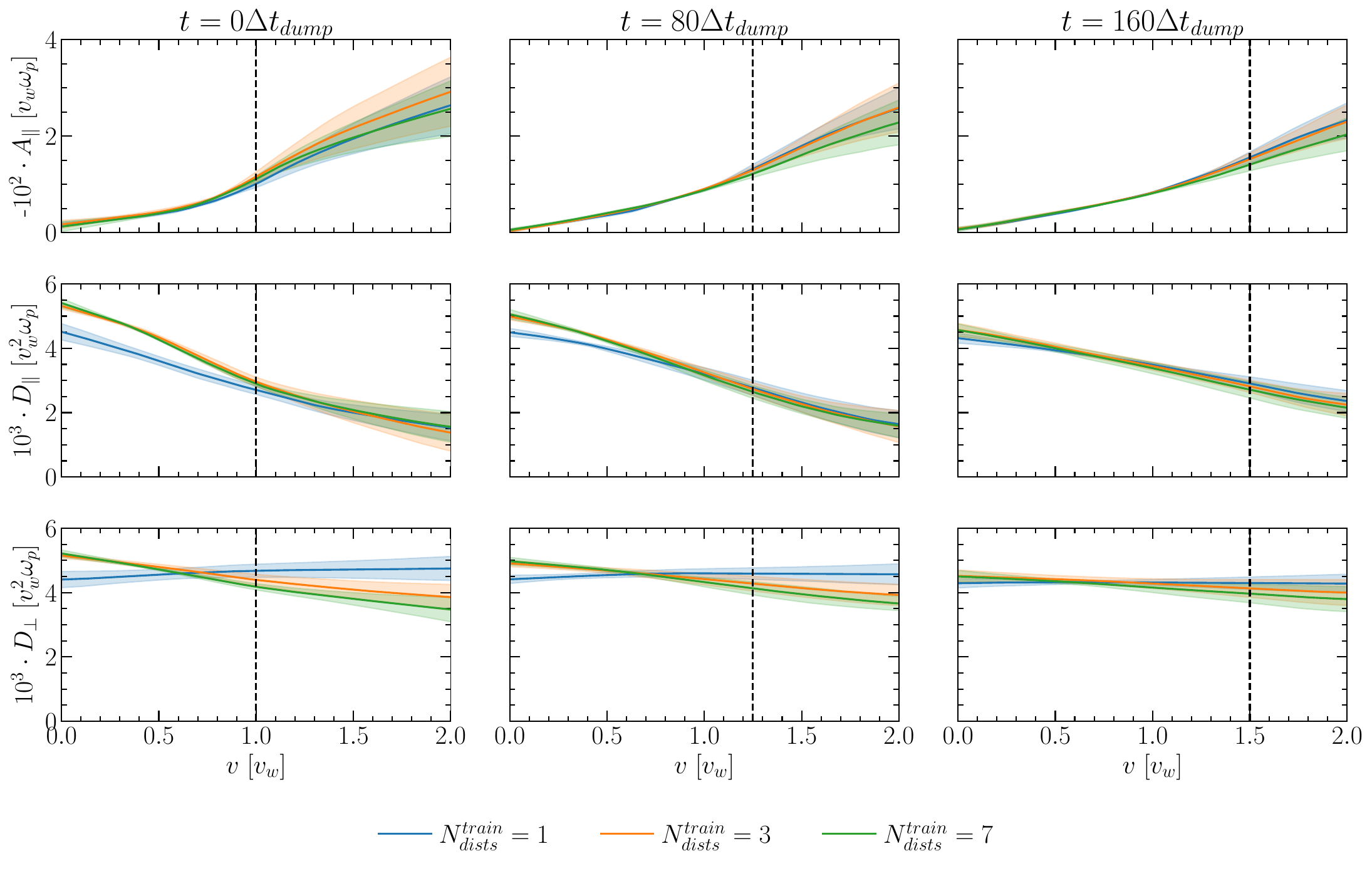}
    \caption{Advection-Diffusion coefficients recovered for PS-NN models using different number of training distributions for Sim-1 (cf. Table~\ref{tab:pic_simulation_parameters}). The full lines represents the mean across 8 initial random seeds. The shadowed regions correspond to the standard deviation. Values for 3 time-steps are shown. Vertical dashed lines represent the maximum velocity across all particles for that simulation time-step. Increasing the number of training distributions decreases the standard deviation across seeds (for regions where data exists) and provides a better estimate of the coefficients. It is clear that using single distribution function (which nonetheless covers the full range of the phase space) is not sufficient to recover the correct coefficients.}
    \label{fig:n_train_dists-AD-sim1-nn}
\end{figure}

In Figure~\ref{fig:n_train_dists-AD-sim1-nn} we plot the advection-diffusion coefficients obtained for $N_{train}^{dists}=1,3,7$ across random seeds for 3 different time-steps (simulation ends at $t\approx187\Delta t_{dump}$). The average vales are shown in full line and the standard deviation as a shaded region. It is clear that $N_{dists}^{train}=1$ produces on average an incorrect estimate of the diffusion coefficients, while differences between $N^{train}_{dists}=3$ and  $N^{train}_{dists}=7$ are not as significant (seem to occur mostly at higher $v$). Furthermore, increasing $N_{dists}^{train}$ clearly reduces the standard deviation of the coefficients across $v$ for regions where data exists (left of the dashed vertical lines). The models are not expected to agree outside regions where no training data is provided.

\subsection{PS-Tensor: Enforcing smoothness via regularization versus $N_t$}
\label{app:regularization_vs_nt}
Regarding the PS-Tensor approach, there exist two clear ways of constraining the operator to recover, what we expect to be, a smooth function. They are: 1) to impose a regularization penalty in the loss function to minimize the norm of its derivatives; 2) reduce the number of degrees of freedom. The latter is already enforced in velocity-space by assuming symmetries in the operator form. To constrain the degrees of freedom over time, the equivalent solution would be to reduce the number of time-steps $N_{t}$ over which the tensor is defined. Similarly, we can impose a regularization penalty over its time-derivatives. These two approaches can of course be combined, but for simplicity we will consider them separately to illustrate the impact they have on the final operator and the corresponding rollout error. Additionally, for this analysis we will use only Sim-1 and $N_{dists}^{train}=7$.

\begin{figure}
    \centering
    \includegraphics[width=0.6\linewidth]{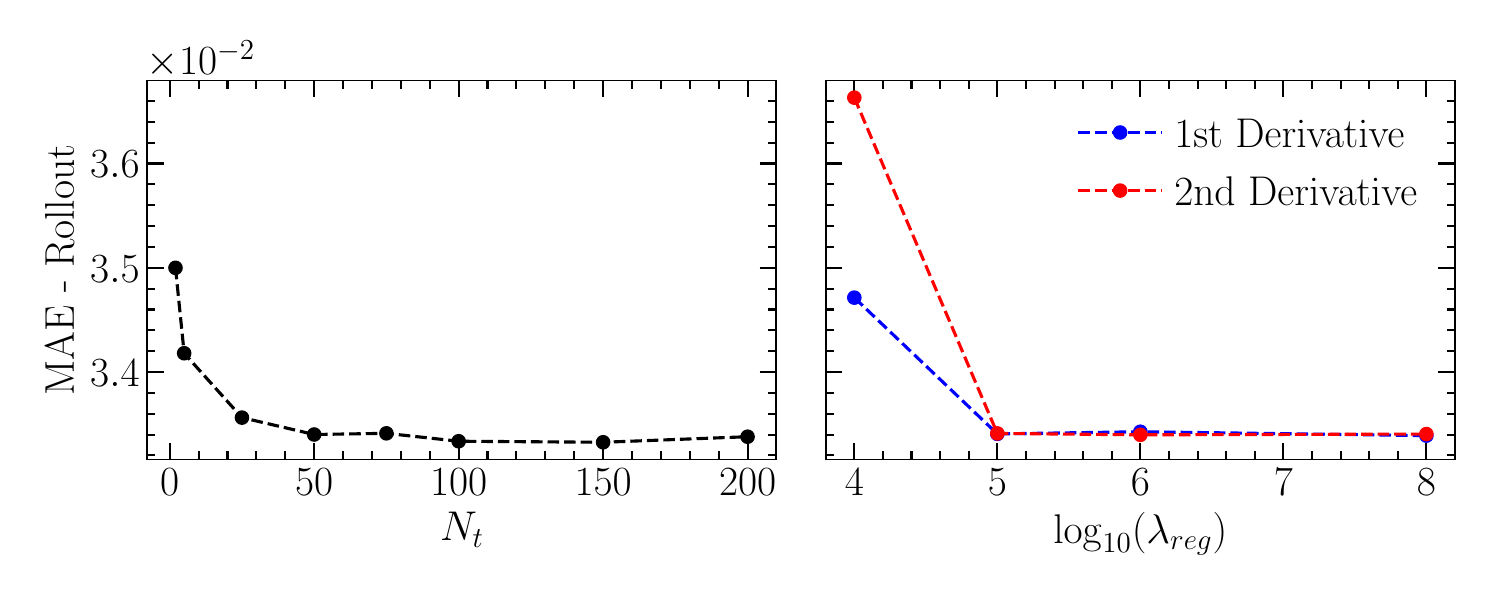}
    \caption{Comparison of rollout error between two different approaches to constrain the time-dependent PS-Tensor collision operators degrees of freedom retrieved for Sim-1 (cf. Table~\ref{tab:pic_simulation_parameters}). We either constrain the number of time-steps $N_t$ over which the tensor is defined (left), or penalize the norm of its time-derivatives (either the 1st or 2nd derivative) via the loss (right). The minimum errors of the different approaches are equivalent.}
    \label{fig:mae-n_t_vs_regularization-tensor}
\end{figure}

In Figure~\ref{fig:mae-n_t_vs_regularization-tensor} we compare the test rollout error of the different approaches. In particular we vary: 1) $N_t$ from $1$ (no time dependence) to $200$ (more free time-steps than time-steps in the simulation); 2) the regularization weight $\lambda_{reg}$ from $10^{-4}$ to $10^{-8}$ (in this case the total loss function is equal to $\mathcal{L} = \mathcal{L}_{data} + \lambda_{reg}\mathcal{L}_{reg}$ with $\mathcal{L}_{reg}=||\partial_t^nA|| + ||\partial_t^nD_\parallel|| + ||\partial_t^nD_\perp||$ and $\mathcal{L}_{data}$ given by~\eqref{eq:loss}).
We observe the expected behaviour for both approaches. If we limit too much the degrees of freedom along the time-axis, either by reducing $N_{t}$ or imposing too much regularization, the performance degrades. Furthermore, the minimum error achieved is similar between the two approaches.

\begin{figure}
    \centering
    \includegraphics[width=0.8\linewidth]{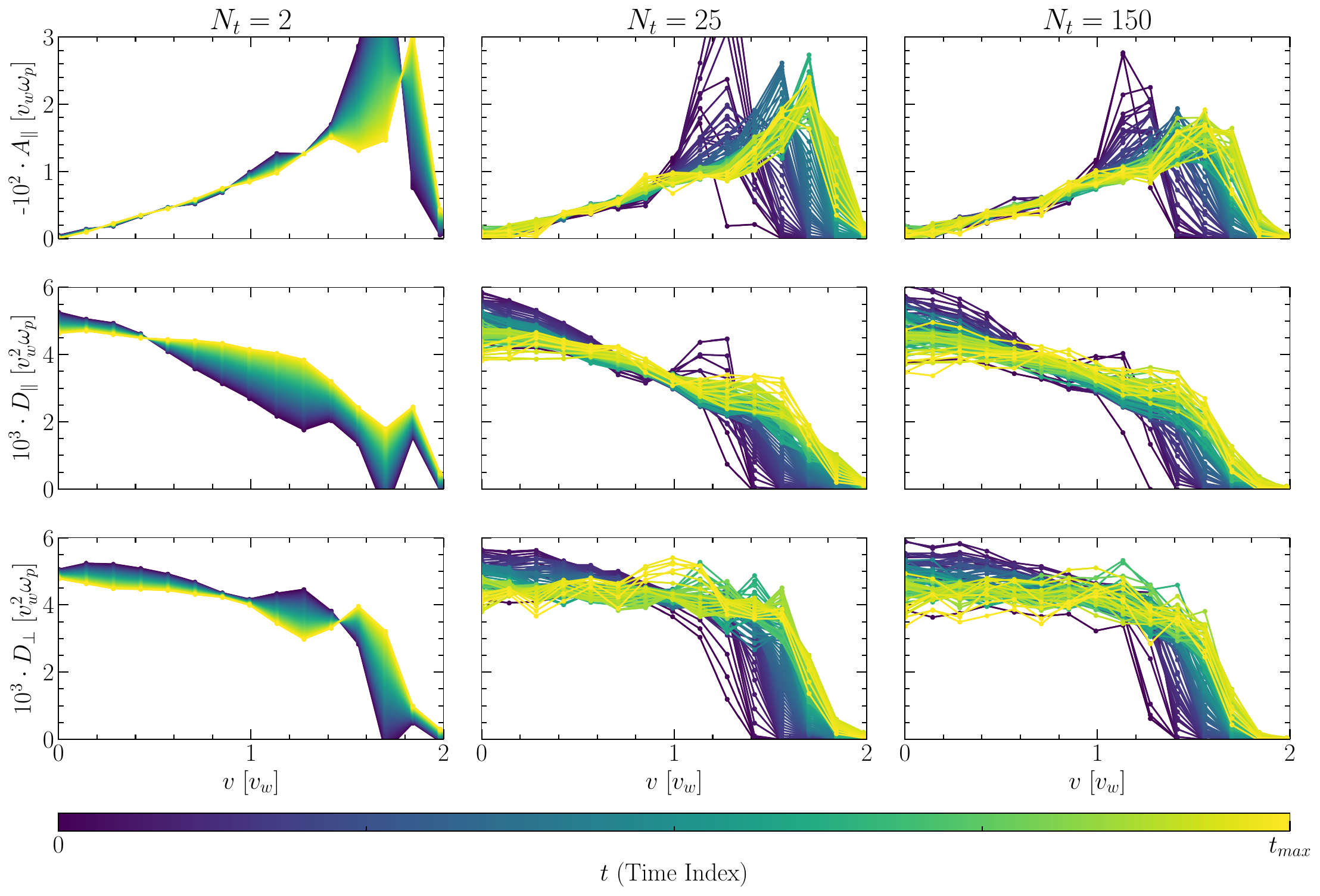}
    \caption{Comparison between operators retrieved for Sim-1 (cf.~Table~\ref{tab:pic_simulation_parameters}) using the PS-Tensor method with different values of $N_{t}$ (higher $N_t$ equals more degrees of freedom). Linear interpolation between neighboring time-steps is used for cases where the learned time-grid is coarser than the time resolution. From the examples shown, the best performing case is $N_{t}=150$ (with, nonetheless, comparable performance to $N_{t}=25$).}
    \label{fig:AD-sim1-comparisons-n_t}
\end{figure}

\begin{figure}
    \centering
    \includegraphics[width=0.8\linewidth]{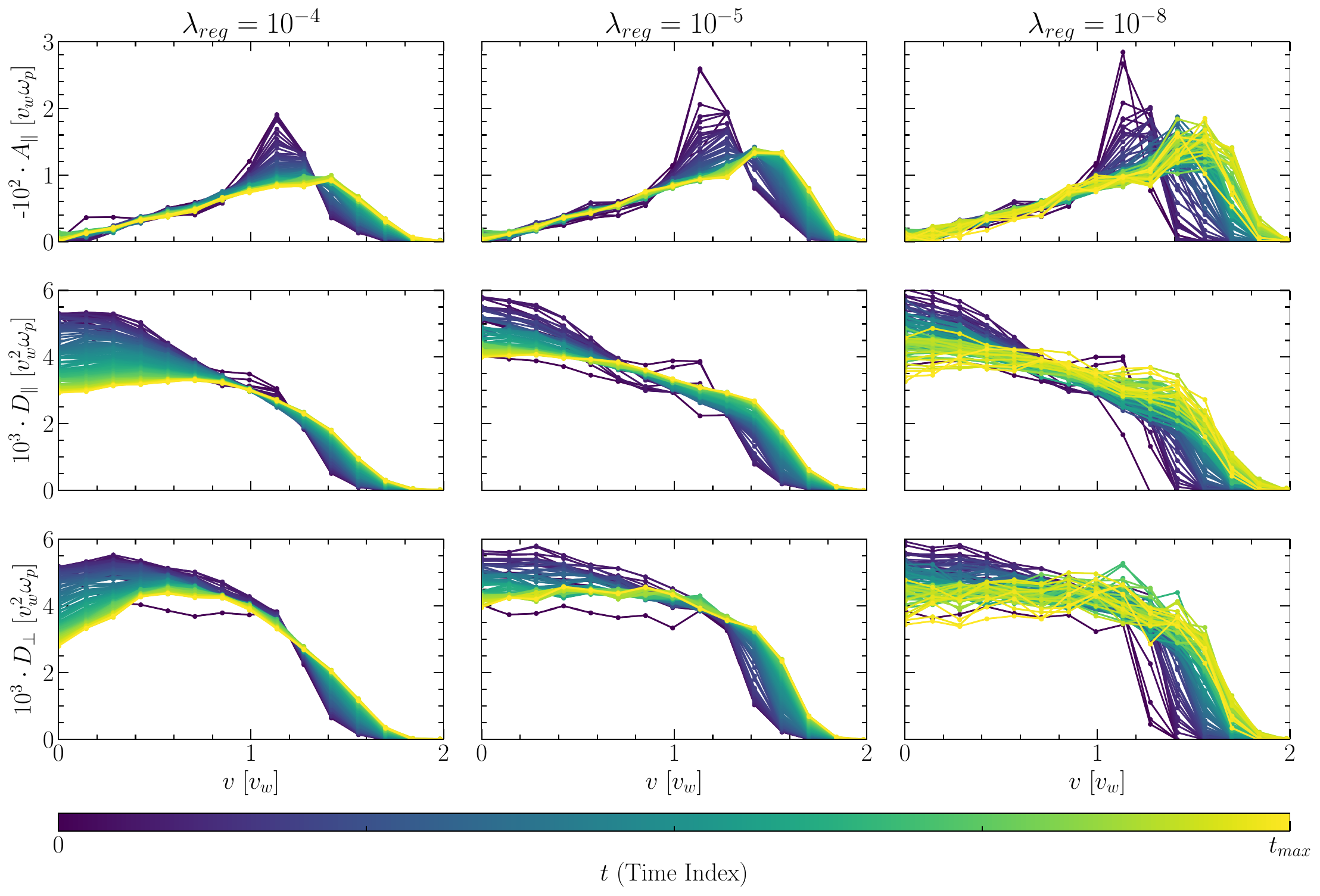}
    \caption{Comparison between operators retrieved  for Sim-1 (cf.~Table~\ref{tab:pic_simulation_parameters}) using the PS-Tensor with different regularization strength applied to its first temporal derivatives. From the examples shown, the best performing cases are $\lambda_{reg}=10^{-5}$ with  $\lambda_{reg}=10^{-8}$ (virtually no regularization). The $\lambda_{reg}=10^{-5}$ case is the model shown in the main body of the paper.}
    \label{fig:AD-sim1-comparisons-regularization-1st}
\end{figure}

\begin{figure}
    \centering
    \includegraphics[width=0.8\linewidth]{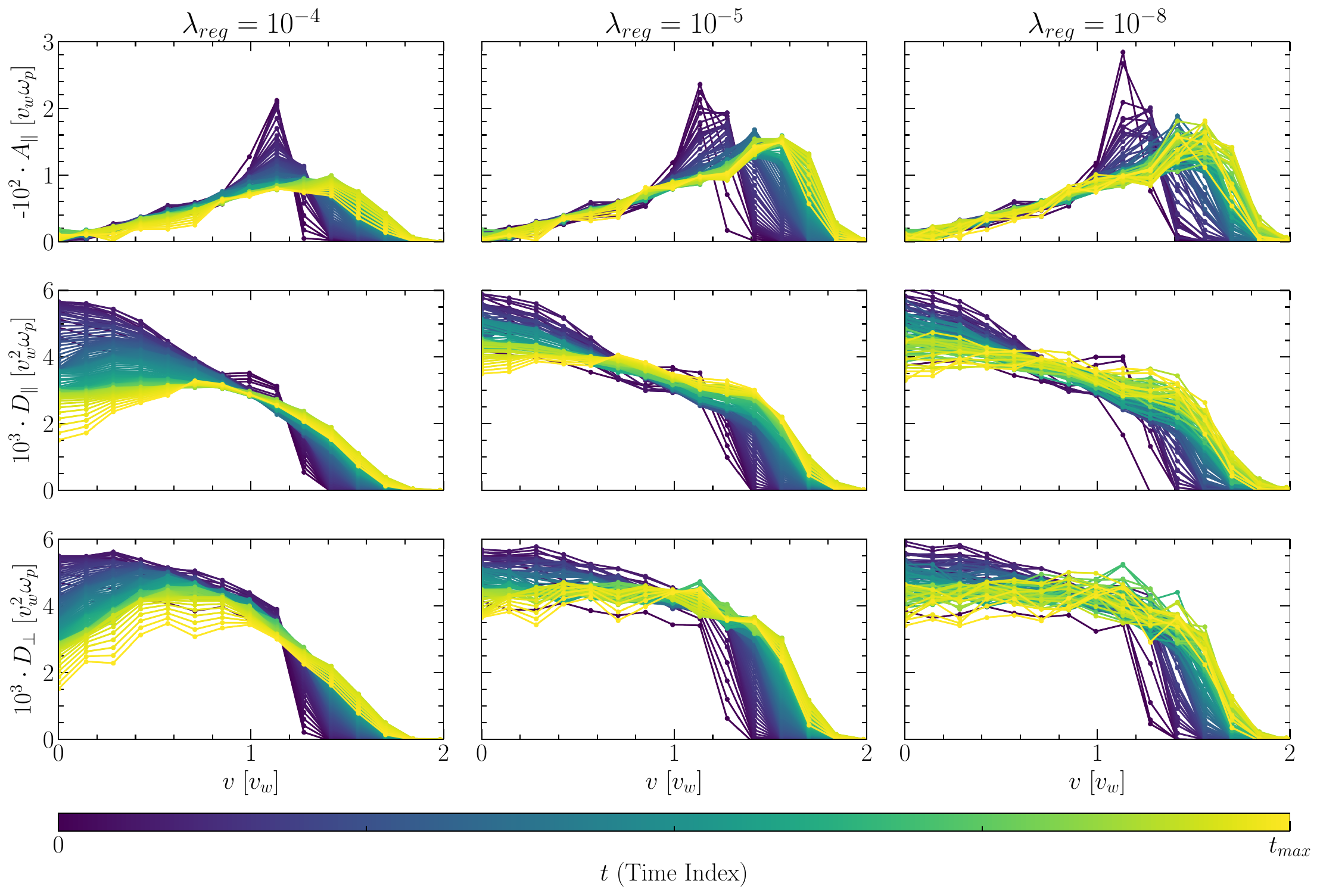}
    \caption{Comparison between operators retrieved for Sim-1 (cf.~Table~\ref{tab:pic_simulation_parameters}) using the PS-Tensor with different regularization strength applied to its second temporal derivatives. From the examples shown, the best performing cases are $\lambda_{reg}=10^{-5}$ with  $\lambda_{reg}=10^{-8}$ (virtually no regularization).}
    \label{fig:AD-sim1-comparisons-regularization-2nd}
\end{figure}

In Figures~\ref{fig:AD-sim1-comparisons-n_t} to~\ref{fig:AD-sim1-comparisons-regularization-2nd} we showcase examples of the operators retrieved for the different strategies.
It is clear that the operators overall values change significantly when stronger restrictions are in place, leading to a decay in performance. Similarly (well) performing operators differ mostly in terms of the noise level, not the average values. The problem of the non-uniqueness of the operator is nonetheless here clearly illustrated, similarly to what was discussed for the NN case in~\ref{app:discussion_non_uniqueness}.

For the main body of the paper we decided to use the model regularized with respect to the first derivative and $\lambda_{reg}=10^{-5}$ (highest value that does not degrade performance) which corresponds to the one in the central column of Figure~\ref{fig:AD-sim1-comparisons-regularization-1st}.

\section{Generalised integro-differential Operator - Extra}
\label{app:K}

\subsection{Importance of the finite-difference scheme}

To clarify the importance of the finite-difference scheme used to compute the gradient operation in~\eqref{eq:K_operator} we provide a similar Pareto-curve analysis to the results shown in Figure~\ref{fig:K-test_l1_avg} now using other possible schemes. The \textit{Forward} case corresponds to the one used in the main body of the paper.
\begin{figure}
    \centering
    \includegraphics[width=0.4\linewidth]{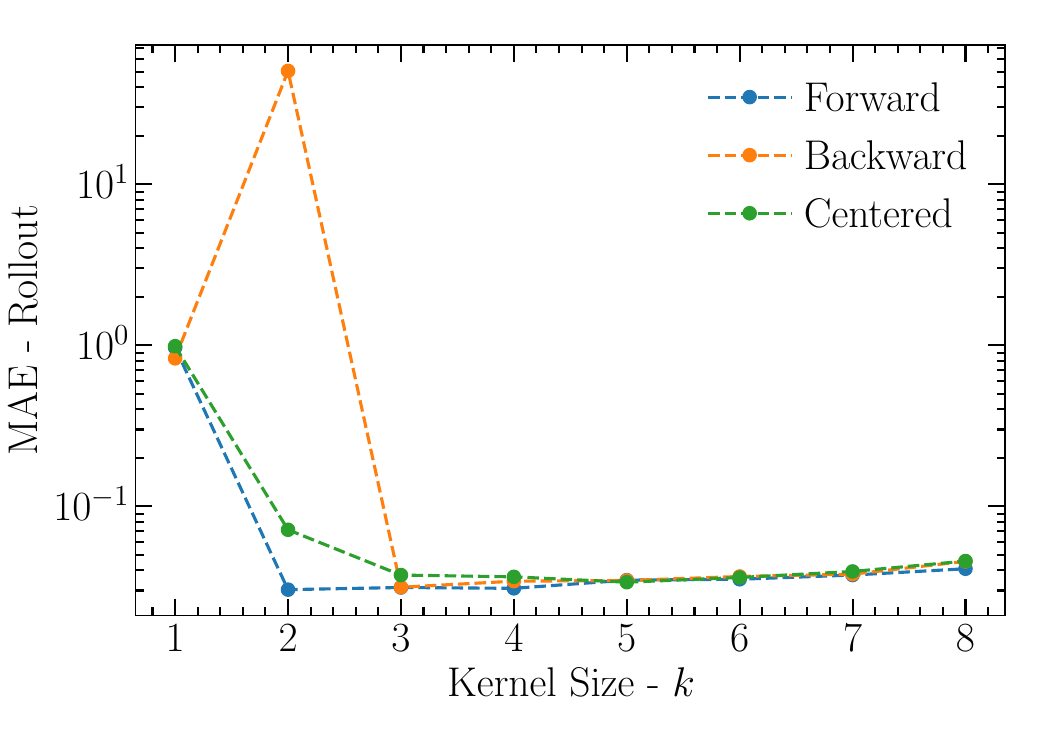}
    \caption{Importance of the choice of the finite difference scheme used to compute outer gradient in \eqref{eq:K_operator}. The forward scheme can perfectly represent both a centred advection and diffusion terms with $k=2$. Therefore, its error plateaus at this value. The backward difference scheme requires $k=3$ to be able to represent the diffusion term since $k<3$ only provides information about $v_x' \leq v_x$. The centred difference scheme suffers from the appearance of spurious terms, which do not make it possible to retrieve a pure advection-diffusion model. These results justify the usage of a forward scheme in the main body of the paper.}
    \label{fig:K-fd_order}
\end{figure}

\begin{figure}
    \centering
    \includegraphics[width=0.7\linewidth]{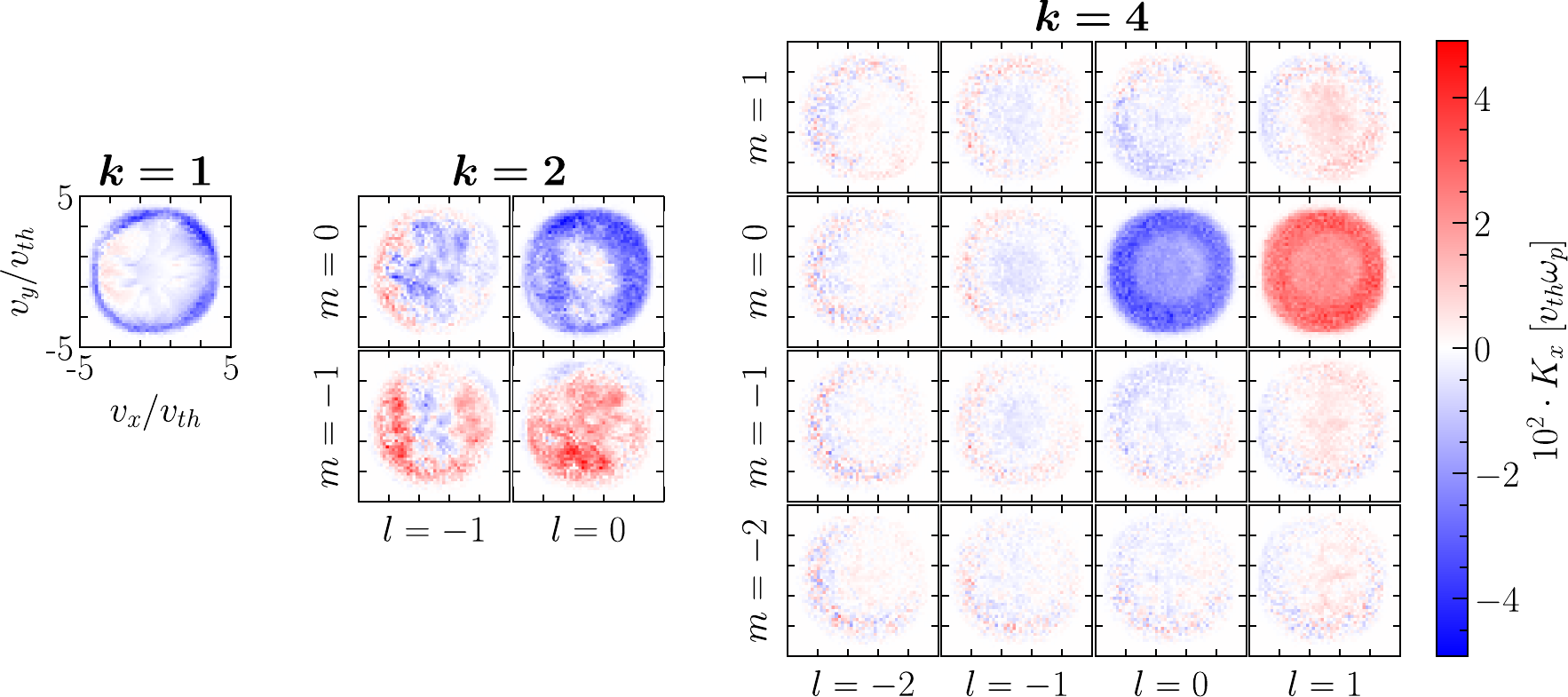}
    \caption{Discrete 4-Dimensional kernel operators recovered for different kernel sizes $k$ using a backward finite-different scheme for the gradient in~\eqref{eq:K_operator}. Unlike for the forward finite-different case shown in Figure~\ref{fig:K_kernel}, $k=2$ is not enough to recover the correct form of the operator (advection-diffusion model). Only $k>3$ allows for such representation. This explains why the error only plateaus for $k>3$ in Figure~\ref{fig:K-fd_order}.}
    \label{fig:K_kernel_1}
\end{figure}

For the \textit{Backward} difference case, the error actually increases when we set $k=2$ and only stabilizes when $k=3$. This is due to the way we define the centering of the convolution kernel which does not allow a diffusion stencil to appear for $k=2$ (see Figure~\ref{fig:K_kernel_1}). Had we defined $n_{lower}=\lfloor k/2 \rfloor$ and $n_{higher}=\lfloor (k-1)/2\rfloor$ (switched with respect to the definition in Section~\ref{methods:K}) the results obtained from the forward and backward case would switch.

\begin{figure}
    \centering
    \includegraphics[width=0.7\linewidth]{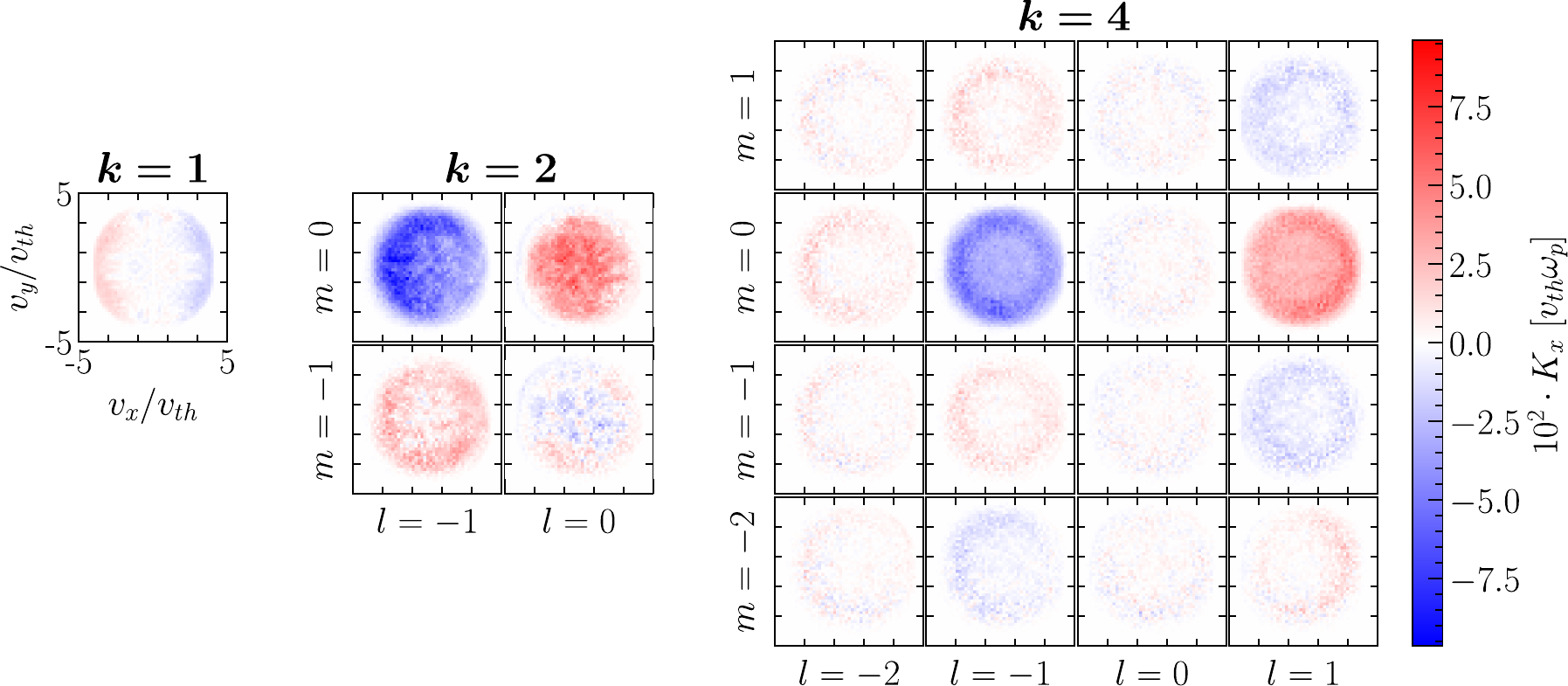}
    \caption{Discrete 4-Dimensional kernel operators recovered for different kernel sizes $k$ using a centered finite-different scheme for the gradient in~\eqref{eq:K_operator}. While $k=2$ should be enough to recover a gradient-like stencil, the centered scheme introduces dependencies on further neighbors, which do not allow the operator to recover an advection-diffusion description. Further increasing $k$ provides the operator more flexibility to remove the spurious contributions but it also increases unnecessarily the number of degrees of freedom. The latter leads to the worse performance observed in Figure~\ref{fig:K-fd_order}.}
    \label{fig:K_kernel_2}
\end{figure}

The reason why the \textit{Centered} scheme does not achieve a minimum error around $k=2$ (neither at $k=3$) is because it can not recover a pure advection-diffusion stencil since extra undesired spurious terms keep appearing in the stencil to allow for the calculation of a second derivative (diffusion) term (see Figure~\ref{fig:K_kernel_2}).

\subsection{Examples of simulation rollouts}

\begin{figure}
    \centering
    \includegraphics[width=0.57\linewidth]{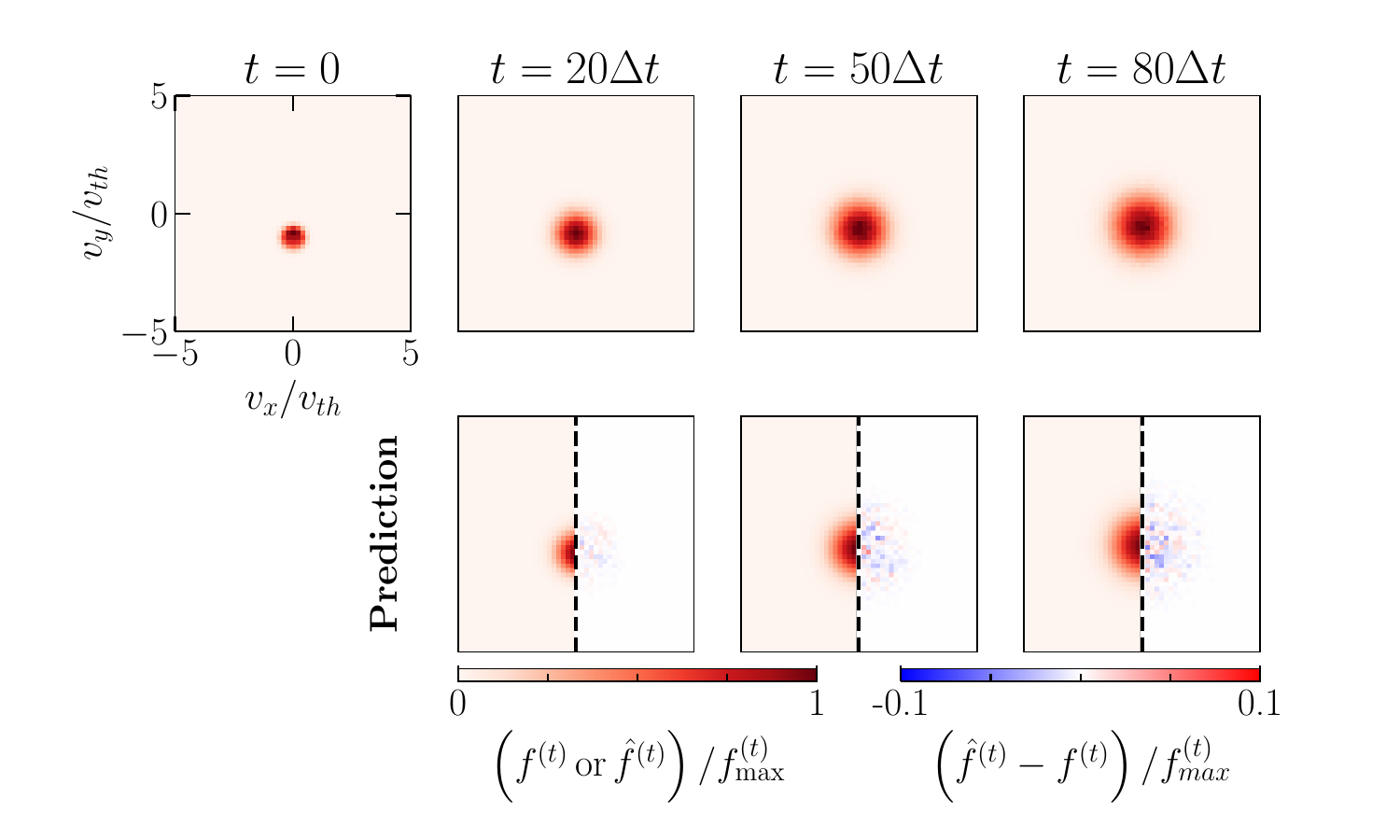} \\
    \includegraphics[width=0.57\linewidth]{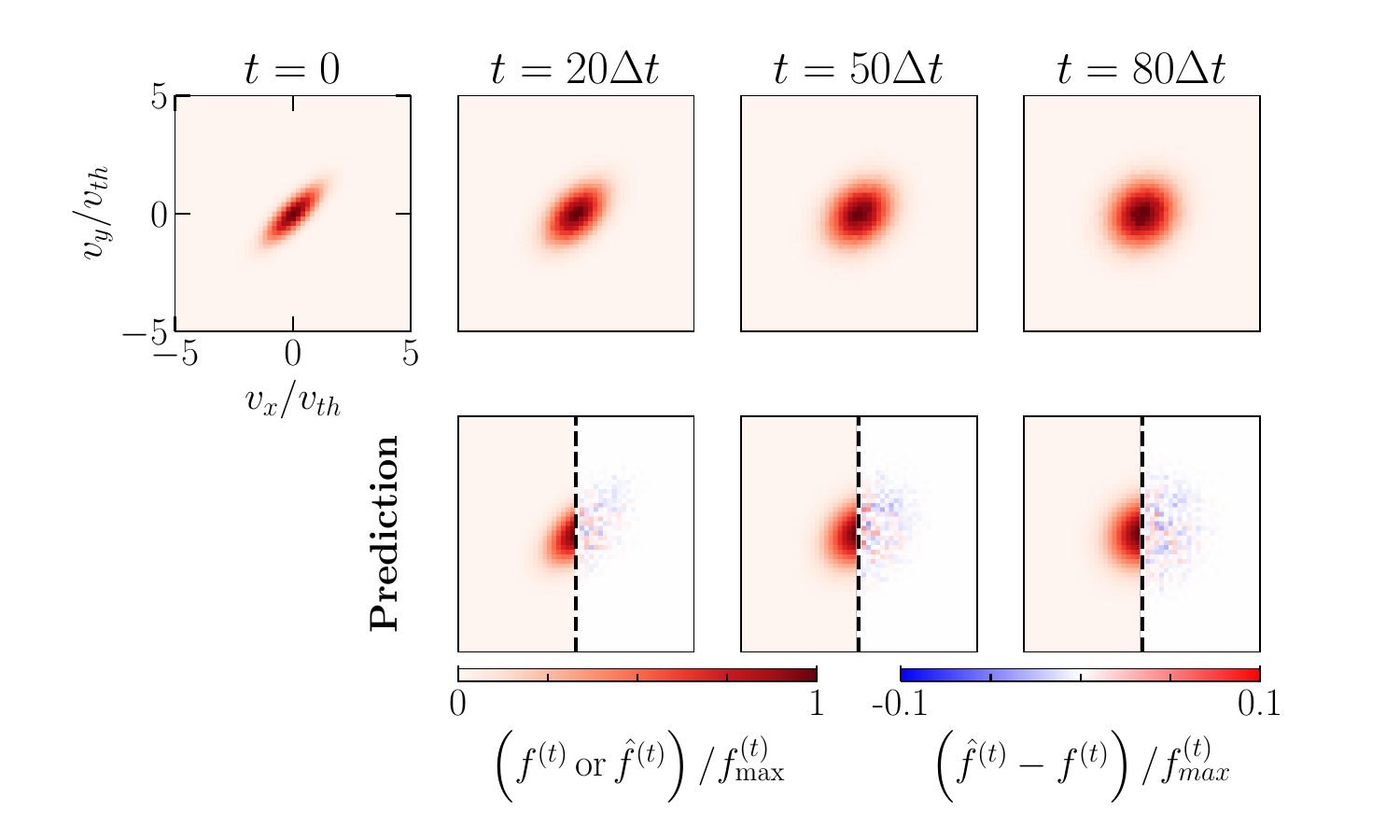} \\
    \includegraphics[width=0.57\linewidth]{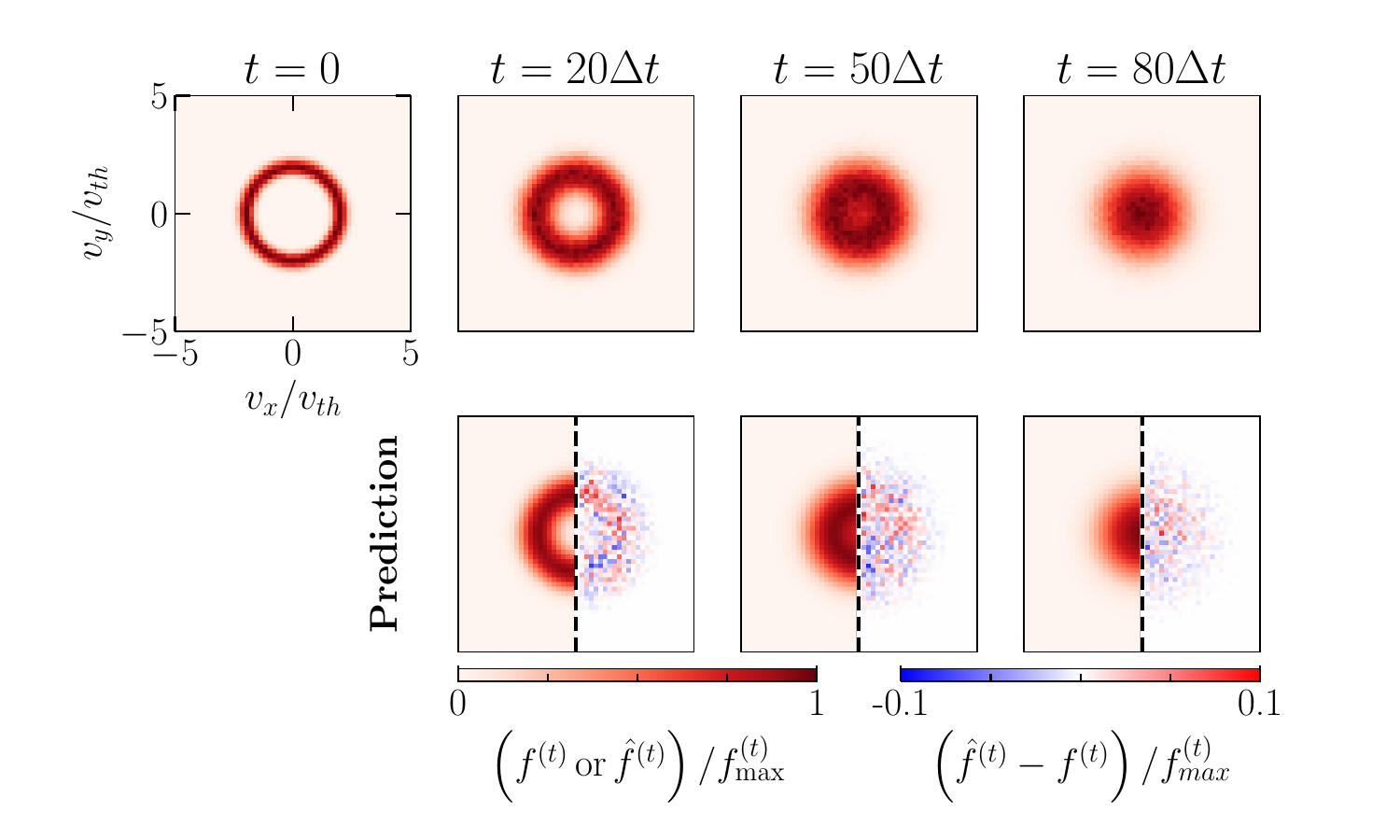}
    \caption{Comparison between the phase space dynamics of subpopulations obtained from the PIC simulation and the predicted evolution using the operator learned in Section~\ref{sec:results_K} ($k=2$). Three different initial subpopulations are shown. As expected, the dynamics are accurately recovered.}
    \label{fig:K-example_rollout}
\end{figure}

In this Appendix we showcase in Figure~\ref{fig:K-example_rollout} a small set of examples of simulation rollouts for different initial subpopulations using the integro-differential operator recovered in Section~\ref{sec:results_K}. Overall we observe that the learned operator can accurately predict the phase space dynamics produced by the PIC simulation further demonstrating the validity of the proposed approach.

\section{Table of Symbols}

In Table~\ref{tab:symbols} we provide an auxiliary summary of the symbols defined throughout the work to facilitate the readability.

\begin{table}[htbp]
\centering
\caption{Summary of Mathematical Symbols and Variables}
\label{tab:symbols}
\begin{tabular}{ll}
\hline
\textbf{Symbol} & \textbf{Description} \\
\hline
\multicolumn{2}{l}{\textbf{Fundamental Quantities \& Functions}} \\
$f$ & Velocity distribution function \\
$\mathcal{C}$ & Collision operator \\
$\boldsymbol{v}$ & Velocity vector in $N$-dimensional space \\
$v_i$ & $i$-th Cartesian component of velocity \\
$v_w$ & Radius of the waterbag velocity distribution \\
$v_{th}$ & Thermal velocity \\
$t$ & Time \\
$n_0$ & Background number electron density \\
$\omega_p$ & Plasma frequency \\
\hline
\multicolumn{2}{l}{\textbf{Fokker-Planck (FP) Operator Terms}} \\
$\mathcal{C}[f]_{FP}$ & Fokker-Planck collision operator \\
$\boldsymbol{A}$ & Advection vector in velocity space \\
$\boldsymbol{D}$ & Diffusion tensor in velocity space \\
$A_i$ & Advection component for the $i$-th velocity component \\
$D_{ij}$ & Diffusion component for the pair of $(i,j)$-th velocity components \\
$A_\parallel$ & Advection component parallel to particle propagation \\
$D_\parallel$ & Diffusion component parallel to particle propagation \\
$D_\perp$ & Diffusion component perpendicular to particle propagation \\
$\mathbf{I}$ & Identity matrix \\
$\hat{\boldsymbol{v}}$ & Unit velocity vector ($\boldsymbol{v}/v$) \\
\hline
\multicolumn{2}{l}{\textbf{Integro-Differential Operator Terms}} \\
$\mathcal{C}[f]_{ID}$ & Integro-differential form of the collision operator \\
$\boldsymbol{K}$ & 5-Dimensional Kernel operator \\
$K_i$ & 4-Dimensional Kernel operator for the $i$-th velocity component \\
$k$ & Size of the discrete kernel (e.g., $k \times k$ cells) \\
$l, m$ & Discrete indices for velocity grid offsets in the kernel \\
\hline
\multicolumn{2}{l}{\textbf{Optimization \& Learning Parameters}} \\
$\hat{f}$ & Predicted distribution function \\
$\boldsymbol{\theta}$ & Free learnable parameters (NN weights or tensor values) \\
$\mathcal{L}$ & Scalar loss function (e.g., Mean Absolute Error) \\
$\mathcal{L}_{N_u}$ & Loss function for a specific unrolling length \\
$\mathcal{L}_{data}$ & Loss function from phase-space error \\
$\mathcal{L}_{reg}$ & Loss function from regularization term \\
$\mathrm{MAE-Rollout}$ & Mean Absolute Error over a long-term rollout \\
$N_v$ & Number of operator grid points in velocity space \\
$N_t$ & Number of grid points in time \\
$N_u$ & Temporal unrolling length (curriculum stage horizon) \\
$N_{train}^{dists}$ & Number of subpopulations used for training \\
$\alpha$ & Learning rate \\
$\lambda_{reg}$ & Regularization weight (penalty strength) \\
\hline
\multicolumn{2}{l}{\textbf{Simulation Parameters}} \\
$\Delta_x$ & Spatial grid resolution in PIC simulations \\
$\Delta_v$ & Phase space (velocity) grid resolution \\
$\Delta t$ & Simulation time-step size \\
$x_{max}$ & Box size (spatial domain limit) \\
$t_{max}$ & Maximum simulation time \\
$t_{dump}$ & Diagnostic dumping period \\
$N_{ppc}$ & Number of particles per cell \\
$N_{dump}$ & Number of diagnostic dumping periods per interval \\
$m$ & Particle shape function order \\
\hline
\end{tabular}
\end{table}

\bibliographystyle{iopart-num}
\bibliography{main.bib}

@article{alves2018efficient,
  title={Efficient nonthermal particle acceleration by the kink instability in relativistic jets},
  author={Alves, E Paulo and Zrake, Jonathan and Fiuza, Frederico},
  journal={Physical review letters},
  volume={121},
  number={24},
  pages={245101},
  year={2018},
  publisher={APS}
}

@inproceedings{ansel2024pytorch,
  title={Pytorch 2: Faster machine learning through dynamic python bytecode transformation and graph compilation},
  author={Ansel, Jason and Yang, Edward and He, Horace and Gimelshein, Natalia and Jain, Animesh and Voznesensky, Michael and Bao, Bin and Bell, Peter and Berard, David and Burovski, Evgeni and others},
  booktitle={Proceedings of the 29th ACM International Conference on Architectural Support for Programming Languages and Operating Systems, Volume 2},
  pages={929--947},
  year={2024}
}

@book{atzeni2004physics,
  title={The Physics of Inertial Fusion: BeamPlasma Interaction, Hydrodynamics, Hot Dense Matter},
  author={Atzeni, Stefano and Meyer-ter-Vehn, J{\"u}rgen},
  number={125},
  year={2004},
  publisher={Oxford University Press}
}

@article{baalrud2012transport,
  title={Transport coefficients in strongly coupled plasmas},
  author={Baalrud, Scott D},
  journal={Physics of Plasmas},
  volume={19},
  number={3},
  year={2012},
  publisher={AIP Publishing}
}

@article{baalrud2014extending,
  title={Extending plasma transport theory to strong coupling through the concept of an effective interaction potential},
  author={Baalrud, Scott D and Daligault, J{\'e}r{\^o}me},
  journal={Physics of Plasmas},
  volume={21},
  number={5},
  year={2014},
  publisher={AIP Publishing}
}

@article{bergeson2025experimental,
  title={Experimental and computational study of phase space dynamics in strongly coupled plasmas with steep density gradients},
  author={Bergeson, Scott and Schlitters, Matthew and Miller, Matthew and Farley, Ben and Sieverts, Devin and Murillo, Michael S and Haack, Jeffrey R},
  journal={Physics of Plasmas},
  volume={32},
  number={3},
  year={2025},
  publisher={AIP Publishing}
}

@article{bhatnagar1954model,
  title={A model for collision processes in gases. I. Small amplitude processes in charged and neutral one-component systems},
  author={Bhatnagar, Prabhu Lal and Gross, Eugene P and Krook, Max},
  journal={Physical Review},
  volume={94},
  number={3},
  pages={511},
  year={1954},
  publisher={APS}
}

@book{birdsall2018plasma,
  title={Plasma physics via computer simulation},
  author={Birdsall, Charles K and Langdon, A Bruce},
  year={2018},
  publisher={CRC press}
}

@incollection{boltzmann1970weitere,
  title={Weitere studien {\"u}ber das w{\"a}rmegleichgewicht unter gasmolek{\"u}len},
  author={Boltzmann, Ludwig},
  booktitle={Kinetische Theorie II: Irreversible Prozesse Einf{\"u}hrung und Originaltexte},
  pages={115--225},
  year={1970},
  publisher={Springer}
}

@article{braams1987differential,
  title={Differential form of the collision integral for a relativistic plasma},
  author={Braams, Bastiaan J and Karney, Charles FF},
  journal={Physical Review Letters},
  volume={59},
  number={16},
  pages={1817},
  year={1987},
  publisher={APS}
}

@article{camporeale2022data,
  title={Data-driven discovery of Fokker-Planck equation for the Earth's radiation belts electrons using Physics-Informed neural networks},
  author={Camporeale, Enrico and Wilkie, George J and Drozdov, Alexander Y and Bortnik, Jacob},
  journal={Journal of Geophysical Research: Space Physics},
  volume={127},
  number={7},
  pages={e2022JA030377},
  year={2022},
  publisher={Wiley Online Library}
}

@article{carvalho2024learning,
  title={Learning the dynamics of a one-dimensional plasma model with graph neural networks},
  author={Carvalho, Diogo D and Ferreira, Diogo R and Silva, Luis O},
  journal={Machine Learning: Science and Technology},
  volume={5},
  number={2},
  pages={025048},
  year={2024},
  publisher={IOP Publishing}
}

@article{carvalho2026learning,
  title={Learning collision operators from plasma phase space data using differentiable simulators},
  author={Carvalho, Diogo D and Bilbao, Pablo J and Mori, Warren B and Silva, Luis O and Alves, E Paulo},
  journal={arXiv preprint arXiv:2601.10885},
  year={2026},
}

@article{chung2023data,
  title={Data-driven stochastic particle scheme for collisional plasma simulations},
  author={Chung, Kyoungseoun and Fei, Fei and Gorji, M Hossein and Jenny, Patrick},
  journal={Journal of Computational Physics},
  volume={492},
  pages={112400},
  year={2023},
  publisher={Elsevier}
}

@article{courant1928partiellen,
  title={{\"U}ber die partiellen Differenzengleichungen der mathematischen Physik},
  author={Courant, Richard and Friedrichs, Kurt and Lewy, Hans},
  journal={Mathematische Annalen},
  volume={100},
  number={1},
  pages={32--74},
  year={1928},
  publisher={Springer}
}

@article{escande2007can,
  title={When can the Fokker-Planck equation describe anomalous or chaotic transport?},
  author={Escande, DF and Sattin, F},
  journal={Physical Review Letters},
  volume={99},
  number={18},
  pages={185005},
  year={2007},
  publisher={APS}
}

@article{esirkepov2001exact,
  title={Exact charge conservation scheme for particle-in-cell simulation with an arbitrary form-factor},
  author={Esirkepov, T Zh},
  journal={Computer Physics Communications},
  volume={135},
  number={2},
  pages={144--153},
  year={2001},
  publisher={Elsevier}
}

@inproceedings{fonseca2002osiris,
  title={OSIRIS: A three-dimensional, fully relativistic particle in cell code for modeling plasma based accelerators},
  author={Fonseca, Ricardo A and Silva, Luis O and Tsung, Frank S and Decyk, Viktor K and Lu, Wei and Ren, Chuang and Mori, Warren B and Deng, Shaogui and Lee, Shiyoun and Katsouleas, T and others},
  booktitle={Computational Science—ICCS 2002: International Conference Amsterdam, The Netherlands, April 21--24, 2002 Proceedings, Part III 2},
  pages={342--351},
  year={2002},
  organization={Springer}
}

@article{gargate2011ion,
  title={Ion acceleration in non-relativistic astrophysical shocks},
  author={Gargat{\'e}, L and Spitkovsky, Anatoly},
  journal={The Astrophysical Journal},
  volume={744},
  number={1},
  pages={67},
  year={2011},
  publisher={IOP Publishing}
}

@article{hockney1971measurements,
  title={Measurements of collision and heating times in a two-dimensional thermal computer plasma},
  author={Hockney, RW},
  journal={Journal of Computational Physics},
  volume={8},
  number={1},
  pages={19--44},
  year={1971},
  publisher={Elsevier}
}

@book{hockney2021computer,
  title={Computer simulation using particles},
  author={Hockney, Roger W and Eastwood, James W},
  year={2021},
  publisher={CRC Press}
}

@article{holloway2021acceleration,
  title={Acceleration of Boltzmann collision integral calculation using machine learning},
  author={Holloway, Ian and Wood, Aihua and Alekseenko, Alexander},
  journal={Mathematics},
  volume={9},
  number={12},
  pages={1384},
  year={2021},
  publisher={MDPI}
}

@article{hwang2020trend,
  title={Trend to equilibrium for the kinetic Fokker-Planck equation via the neural network approach},
  author={Hwang, Hyung Ju and Jang, Jin Woo and Jo, Hyeontae and Lee, Jae Yong},
  journal={Journal of Computational Physics},
  volume={419},
  pages={109665},
  year={2020},
  publisher={Elsevier}
}

@article{isliker2017particle,
  title={Particle acceleration and fractional transport in turbulent reconnection},
  author={Isliker, Heinz and Pisokas, Theophilos and Vlahos, Loukas and Anastasiadis, Anastasios},
  journal={The Astrophysical Journal},
  volume={849},
  number={1},
  pages={35},
  year={2017},
  publisher={IOP Publishing}
}

@article{isliker2017fractional,
  title = {Fractional Transport in Strongly Turbulent Plasmas},
  author = {Isliker, Heinz and Vlahos, Loukas and Constantinescu, Dana},
  journal = {Phys. Rev. Lett.},
  volume = {119},
  issue = {4},
  pages = {045101},
  numpages = {5},
  year = {2017},
  publisher = {American Physical Society},
}

@article{kawaguchi2022physics,
  title={Physics-informed neural networks for solving the Boltzmann equation of the electron velocity distribution function in weakly ionized plasmas},
  author={Kawaguchi, Satoru and Murakami, Tomoyuki},
  journal={Japanese Journal of Applied Physics},
  volume={61},
  number={8},
  pages={086002},
  year={2022},
  publisher={IOP Publishing}
}

@article{kingma2014adam,
  title={Adam: A method for stochastic optimization},
  author={Kingma, Diederik P},
  journal={arXiv preprint arXiv:1412.6980},
  year={2014}
}

@article{landau1936kinetische,
  title={Die kinetische Gleichung fuer den fall Coulombscher wechselwirkung},
  author={Landau, Lev D},
  journal={Phys. Z. Sowjet. 10 (1936)},
  volume={10},
  pages={163},
  year={1936}
}

@inproceedings{langdon1970nonphysical,
  title={Nonphysical modifications to oscillations, fluctuations, and collisions due to space-time differencing},
  author={Langdon, A Bruce},
  booktitle={Proceedings of the fourth conference on numerical simulation of plasmas},
  pages={467--495},
  year={1970}
}

@article{langdon1970theory,
  title={Theory of Plasma Simulation Using Finite-Size Particles},
  author={Langdon, A Bruce and Birdsall, Charles K},
  journal={The Physics of Fluids},
  volume={13},
  number={8},
  pages={2115--2122},
  year={1970},
  publisher={AIP Publishing}
}

@article{lee2021model,
  title={The model reduction of the Vlasov--Poisson--Fokker--Planck system to the Poisson--Nernst--Planck system via the deep neural network approach},
  author={Lee, Jae Yong and Jang, Jin Woo and Hwang, Hyung Ju},
  journal={ESAIM: Mathematical Modelling and Numerical Analysis},
  volume={55},
  number={5},
  pages={1803--1846},
  year={2021},
  publisher={EDP Sciences}
}

@article{lee2023oppinn,
  title={opPINN: Physics-informed neural network with operator learning to approximate solutions to the Fokker-Planck-Landau equation},
  author={Lee, Jae Yong and Jang, Juhi and Hwang, Hyung Ju},
  journal={Journal of Computational Physics},
  volume={480},
  pages={112031},
  year={2023},
  publisher={Elsevier}
}

@article{lemoine2023particle,
  title={Particle transport through localized interactions with sharp magnetic field bends in MHD turbulence},
  author={Lemoine, Martin},
  journal={Journal of Plasma Physics},
  volume={89},
  number={5},
  pages={175890501},
  year={2023},
  publisher={Cambridge University Press}
}

@article{lemoine2022first,
  title={First-principles fermi acceleration in magnetized turbulence},
  author={Lemoine, Martin},
  journal={Physical Review Letters},
  volume={129},
  number={21},
  pages={215101},
  year={2022},
  publisher={APS}
}

@article{li2024solving,
  title={Solving the boltzmann equation with a neural sparse representation},
  author={Li, Zhengyi and Wang, Yanli and Liu, Hongsheng and Wang, Zidong and Dong, Bin},
  journal={SIAM Journal on Scientific Computing},
  volume={46},
  number={2},
  pages={C186--C215},
  year={2024},
  publisher={SIAM}
}

@article{lou2021physics,
  title={Physics-informed neural networks for solving forward and inverse flow problems via the Boltzmann-BGK formulation},
  author={Lou, Qin and Meng, Xuhui and Karniadakis, George Em},
  journal={Journal of Computational Physics},
  volume={447},
  pages={110676},
  year={2021},
  publisher={Elsevier}
}

@article{martins2009ion,
  title={Ion dynamics and acceleration in relativistic shocks},
  author={Martins, SF and Fonseca, RA and Silva, LO and Mori, Warren B},
  journal={The Astrophysical Journal},
  volume={695},
  number={2},
  pages={L189},
  year={2009},
  publisher={IOP Publishing}
}

@article{mcdevitt2023physics,
  title={A physics-informed deep learning model of the hot tail runaway electron seed},
  author={McDevitt, Christopher J},
  journal={Physics of Plasmas},
  volume={30},
  number={9},
  year={2023},
  publisher={AIP Publishing}
}

@article{mcdevitt2025physics,
  title={A physics-constrained deep learning treatment of runaway electron dynamics},
  author={McDevitt, Christopher J and Arnaud, Jonathan S and Tang, Xian-Zhu},
  journal={Physics of Plasmas},
  volume={32},
  number={4},
  year={2025},
  publisher={AIP Publishing}
}

@article{mcdevitt2025efficient,
  title={An efficient surrogate model of secondary electron formation and evolution},
  author={McDevitt, Christopher J and Arnaud, Jonathan S and Tang, Xian-Zhu},
  journal={Physics of Plasmas},
  volume={32},
  number={4},
  year={2025},
  publisher={AIP Publishing}
}

@article{miller2021encoder,
  title={Encoder--decoder neural network for solving the nonlinear Fokker--Planck--Landau collision operator in XGC},
  author={Miller, Marco Andres and Churchill, Randy Michael and Dener, Alp and Chang, Choong-Seock and Munson, Todd and Hager, Robert},
  journal={Journal of Plasma Physics},
  volume={87},
  number={2},
  pages={905870211},
  year={2021},
  publisher={Cambridge University Press}
}

@article{miller2022neural,
  title={Neural-network based collision operators for the Boltzmann equation},
  author={Miller, Sean T and Roberts, Nathan V and Bond, Stephen D and Cyr, Eric C},
  journal={Journal of Computational Physics},
  volume={470},
  pages={111541},
  year={2022},
  publisher={Elsevier}
}

@article{noh2025fpl,
  title={FPL-net: A deep learning framework for solving the nonlinear Fokker--Planck--Landau collision operator for anisotropic temperature relaxation},
  author={Noh, Hyeongjun and Lee, Jimin and Yoon, Eisung},
  journal={Journal of Computational Physics},
  volume={523},
  pages={113665},
  year={2025},
  publisher={Elsevier}
}

@article{oh2025separable,
author = {Oh, Jaemin and Cho, Seung Yeon and Yun, Seok-Bae and Park, Eunbyung and Hong, Youngjoon},
title = {Separable Physics-Informed Neural Networks for Solving the BGK Model of the Boltzmann Equation},
journal = {SIAM Journal on Scientific Computing},
volume = {47},
number = {2},
pages = {C451-C474},
year = {2025},
}

@article{pike2016transport,
  title={Transport coefficients of a relativistic plasma},
  author={Pike, OJ and Rose, SJ},
  journal={Physical Review E},
  volume={93},
  number={5},
  pages={053208},
  year={2016},
  publisher={APS}
}

@article{raissi2019physics,
  title={Physics-informed neural networks: A deep learning framework for solving forward and inverse problems involving nonlinear partial differential equations},
  author={Raissi, Maziar and Perdikaris, Paris and Karniadakis, George E},
  journal={Journal of Computational Physics},
  volume={378},
  pages={686--707},
  year={2019},
  publisher={Elsevier}
}

@article{rosenbluth1957fokker,
  title={Fokker-Planck equation for an inverse-square force},
  author={Rosenbluth, Marshall N and MacDonald, William M and Judd, David L},
  journal={Physical Review},
  volume={107},
  number={1},
  pages={1},
  year={1957},
  publisher={APS}
}

@article{touati2022kinetic,
  title={Kinetic theory of particle-in-cell simulation plasma and the ensemble averaging technique},
  author={Touati, Micha{\"e}l and Codur, Romain and Tsung, Frank and Decyk, Viktor K and Mori, Warren B and Silva, Luis O},
  journal={Plasma Physics and Controlled Fusion},
  volume={64},
  number={11},
  pages={115014},
  year={2022},
  publisher={IOP Publishing}
}

@article{wong2020first,
  title={First-principles demonstration of diffusive-advective particle acceleration in kinetic simulations of relativistic plasma turbulence},
  author={Wong, Kai and Zhdankin, Vladimir and Uzdensky, Dmitri A and Werner, Gregory R and Begelman, Mitchell C},
  journal={The Astrophysical Journal Letters},
  volume={893},
  number={1},
  pages={L7},
  year={2020},
  publisher={IOP Publishing}
}

@article{wong2025energy,
  title={Energy diffusion and advection coefficients in kinetic simulations of relativistic plasma turbulence},
  author={Wong, Kai W and Zhdankin, Vladimir and Uzdensky, Dmitri A and Werner, Gregory R and Begelman, Mitchell C},
  journal={arXiv preprint arXiv:2502.03042},
  year={2025}
}

@article{xiao2021using,
  title={Using neural networks to accelerate the solution of the Boltzmann equation},
  author={Xiao, Tianbai and Frank, Martin},
  journal={Journal of Computational Physics},
  volume={443},
  pages={110521},
  year={2021},
  publisher={Elsevier}
}

@article{xiao2023relaxnet,
  title={RelaxNet: A structure-preserving neural network to approximate the Boltzmann collision operator},
  author={Xiao, Tianbai and Frank, Martin},
  journal={Journal of Computational Physics},
  volume={490},
  pages={112317},
  year={2023},
  publisher={Elsevier}
}

@article{yee1966numerical,
  title={Numerical solution of initial boundary value problems involving Maxwell's equations in isotropic media},
  author={Yee, Kane},
  journal={IEEE Transactions on antennas and propagation},
  volume={14},
  number={3},
  pages={302--307},
  year={1966},
  publisher={Ieee}
}

@article{yu2025physics,
  title={Physics-tailored machine learning reveals unexpected physics in dusty plasmas},
  author={Yu, Wentao and Abdelaleem, Eslam and Nemenman, Ilya and Burton, Justin C},
  journal={Proceedings of the National Academy of Sciences},
  volume={122},
  number={31},
  pages={e2505725122},
  year={2025},
  publisher={National Academy of Sciences}
}

@article{zhao2025data,
  title={Data-driven construction of a generalized kinetic collision operator from molecular dynamics},
  author={Zhao, Yue and Burby, Joshua and Christlieb, Andrew and Lei, Huan},
  journal={Physical Review Letters},
  volume={135},
  number={18},
  pages={185101},
  year={2025},
  publisher={APS}
}

@article{zhao2025fast,
  title={Fast spectral separation method for kinetic equation with anisotropic non-stationary collision operator retaining micro-model fidelity},
  author={Zhao, Yue and Lei, Huan},
  journal={arXiv preprint arXiv:2510.15093},
  year={2025}
}

@article{zhong2022low,
  title={Low-temperature plasma simulation based on physics-informed neural networks: frameworks and preliminary applications},
  author={Zhong, Linlin and Wu, Bingyu and Wang, Yifan},
  journal={Physics of Fluids},
  volume={34},
  number={8},
  year={2022},
  publisher={AIP Publishing}
}

\end{document}